\documentclass[a4paper,12pt]{article}
\usepackage[latin1]{inputenc}
\usepackage{amsmath,amssymb,amsfonts}
\usepackage[authoryear]{natbib}
\usepackage{graphicx}
\usepackage{fullpage}

\title{Bivariate gamma-geometric law and its induced\\ L\'evy process\\
\vspace{.4cm}
\small{\bf{Published in Journal of Multivariate Analysis, volume 109, August 2012,\\ pages 130-145,
DOI: 10.1016/j.jmva.2012.03.004}}}
\author{Wagner Barreto-Souza\vspace{.2cm}\\
\small{Departamento de Estat\'istica, Universidade de S\~ao Paulo}\\
\small{Rua do Mat\~ao, 1010, S\~ao Paulo/SP 05508-090, Brazil}\\
\small{E-mail: wagnerbs85@gmail.com}
}

\begin{document}
\maketitle

\begin{abstract}
 
In this article we introduce a three-parameter extension of the bivariate ex\-po\-nen\-tial-geometric (BEG) law (Kozubowski and Panorska, 2005). We refer to this new distribution as bivariate gamma-geometric (BGG) law. A bivariate random vector $(X,N)$ follows BGG law if $N$ has geometric distribution and $X$ may be represented (in law) as a sum of $N$ independent and identically distributed gamma variables, where these variables are independent of $N$. Statistical properties such as moment generation and characteristic functions, moments and variance-covariance matrix are provided.  The marginal and conditional laws are also studied. We show that BBG distribution is infinitely divisible, just as BEG model is. Further, we provide alternative representations for the BGG distribution and show that it enjoys a geometric stability property. Maximum likelihood estimation and inference are discussed and a reparametrization is proposed in order to obtain orthogonality of the parameters. We present an application to the real data set where our model provides a better fit than BEG model. Our bivariate distribution induces a bivariate L\'evy process with correlated gamma and negative binomial processes, which extends the bivariate L\'evy motion proposed by Kozubowski et al. (2008). The marginals of our L\'evy motion are mixture of gamma and negative binomial processes and we named it $\mbox{BMixGNB}$ motion. Basic properties such as stochastic self-similarity and covariance matrix of the process are presented. The bivariate distribution at fixed time of our $\mbox{BMixGNB}$ process is also studied and some results are derived, including a discussion about maximum likelihood estimation and inference.\\ 

\end{abstract}

\section{Introduction}

Mixed univariate distributions have been introduced and studied in the last years by compounding continuous and discrete distributions. Marshall and Olkin (1997) introduced a class of distributions which can be obtained by minimum and maximum of independent and identically distributed (iid) continuous random variables (independent of the random sample size), where the sample size follows geometric distribution. 

Chahkandi and Ganjali (2009) introduced some lifetime distributions by compounding exponential and power series distributions; this models are called exponential power series (EPS) distributions. Recently, Morais and Barreto-Souza (2011) introduced a class of distributions obtained by mixing Weibull and power series distributions and studied several of its statistical properties. This class contains the EPS distributions and other lifetime models studied recently, for example, the Weibull-geometric distribution (Marshall and Olkin, 1997; Barreto-Souza et al., 2011). The reader is referred to introduction from Morais and Barreto-Souza's (2011) article for a brief literature review about some univariate distributions obtained by compounding.

A mixed bivariate law with exponential and geometric marginals was introduced by Kozubowski and Panorska (2005), and named bivariate exponential-geometric (BEG) distribution. A bivariate random vector $(X,N)$ follows BEG law if admits the stochastic representation: 
\begin{eqnarray}\label{rep}
\left(X,N\right)\stackrel{d}{=}\left(\sum_{i=1}^NX_i,N\right),
\end{eqnarray}
where the variable $N$ follows geometric distribution and $\{X_i\}_{i=1}^\infty$ is a sequence of iid exponential variables, independent of $N$. The BEG law is infinitely divisible and therefore leads a bivariate L\'evy process, in this case, with gamma and negative binomial marginal processes. This bivariate process, named BGNB motion, was introduced and studied by Kozubowski et al. (2008).

Other multivariate distributions involving exponential and geometric distributions have been studied in the literature. Kozubowski and Panorska (2008) introduced and studied a bivariate distribution involving geometric maximum of exponential variables. A trivariate distribution involving geometric sums and maximum of exponentials variables was also recently introduced by Kozubowski et al. (2011).
  
Our chief goal in this article is to introduce a three-parameter extension of the BEG law. We refer to this new three-parameter distribution as bivariate gamma-geometric (BGG) law. Further, we show that this extended distribution is infinitely divisible, and, therefore, it induces a bivariate L\'evy process which has the BGNB motion as particular case. The additional parameter controls the shape of the continuous part of our models. 

Our bivariate distribution may be applied in areas such as hydrology and finance. We here focus in finance applications and use the BGG law for modeling log-returns (the $X_i$'s) corresponding to a daily exchange rate. More specifically, we are interested in modeling cumulative log-returns (the $X$) in growth periods of the exchange rates. In this case $N$ represents the duration of the growth period, where the consecutive log-returns are positive. As mentioned by Kozubowski and Panorska (2005), the geometric sum represented by $X$ in (\ref{rep}) is very useful in several fields including water resources, climate research and finance. We refer the reader to the introduction from Kozubowski and Panorska's (2005) article for a good discussion on practical situations where the random vectors with description (\ref{rep}) may be useful.

The present article is organized as follows. In the Section 2 we introduce the bivariate gamma-geometric law and derive basic statistical properties, including a study of some properties of its marginal and conditional distributions. Further, we show that our proposed law is infinitely divisible. Estimation by maximum likelihood and inference for large sample are addressed in the Section 3, which also contains a proposed reparametrization of the model in order to obtain orthogonality of the parameter in the sense of Cox and Reid (1987). An application to a real data set is presented in the Section 4. The induced L\'evy process is approached in the Section 5 and some of its basic properties are shown. We include a study of the bivariate distribution of the process at fixed time and also discuss estimation of the parameters and inferential aspects. We close the article with concluding remarks in the Section 6.

\section{The law and basic properties}

The bivariate gamma-geometric (BGG) law is defined by the stochastic representation (\ref{rep}) and assuming that $\{X_i\}_{i=1}^\infty$ is a sequence of iid gamma variables independent of $N$ and with probability density function given by $g(x;\alpha,\beta)=\beta^\alpha/\Gamma(\alpha)x^{\alpha-1}e^{-\beta x}$, for $x>0$ and $\alpha,\beta>0$; we denote $X_i\sim\Gamma(\alpha,\beta)$. As before, $N$ is a geometric variable with probability mass function given by $P(N=n)=p(1-p)^{n-1}$, for $n\in\mathbb{N}$; denote $N\sim\mbox{Geom}(p)$. Clearly, the BGG law contains the BEG law as particular case, for the choice $\alpha=1$. 
The joint density function $f_{X,N}(\cdot,\cdot)$ of $(X,N)$ is given by
\begin{eqnarray}\label{density}
f_{X,N}(x,n)=\frac{\beta^{n\alpha}}{\Gamma(\alpha n)}x^{n\alpha-1}e^{-\beta x}p(1-p)^{n-1}, \quad x>0,\,\, n\in\mathbb{N}.
\end{eqnarray} 

Hence, it follows that the joint cumulative distribution function (cdf) of the BGG distribution can be expressed by
\begin{eqnarray*}
P(X\leq x, N\leq n)=p\sum_{j=1}^n(1-p)^{j-1}\frac{\Gamma_{\beta x}(j\alpha)}{\Gamma(j\alpha)}, 
\end{eqnarray*}
for $x>0$ and $n\in\mathbb{N}$, where $\Gamma_x(\alpha)=\int_0^xt^{\alpha-1}e^{-t}dt$ is the incomplete gamma function. 
We will denote $(X,N)\sim\mbox{BGG}(\beta,\alpha,p)$. We now show that $(pX,pN)\stackrel{d}{\rightarrow}(\alpha Z/\beta,Z)$ as $p\rightarrow0^+$, where `$\stackrel{d}{\rightarrow}$' denotes convergence in distribution and $Z$ is a exponential variable with mean 1; for $\alpha=1$, we obtain the result given in the proposition 2.3 from Kozubowski and Panorska (2005). For this, we use the moment generation function of the BGG distribution, which is given in the Subsection 2.2. Hence, we have that
$E(e^{tpX+spN})=\varphi(pt,ps)$, where $\varphi(\cdot,\cdot)$ is given by (\ref{mgf}). Using L'H\^opital's rule, one may check that $E(e^{tpX+spN})\rightarrow(1-s-\alpha t/\beta)^{-1}$ as $p\rightarrow0^+$, which is the moment generation function of $(\alpha Z/\beta,Z)$.

\subsection{Marginal and conditional distributions}

\begin{figure}[h!]
	\centering
		\includegraphics[width=0.33\textwidth]{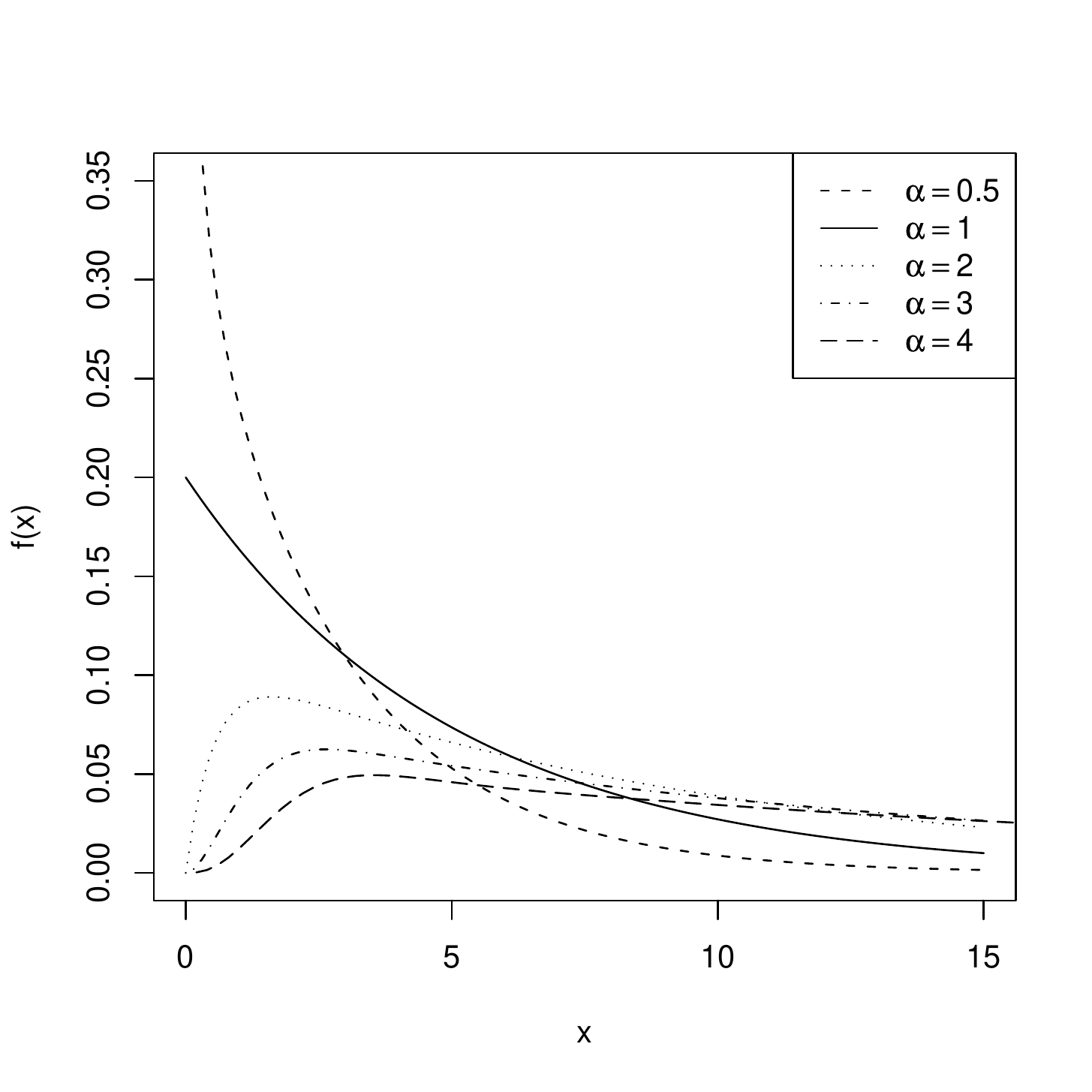}\includegraphics[width=0.33\textwidth]{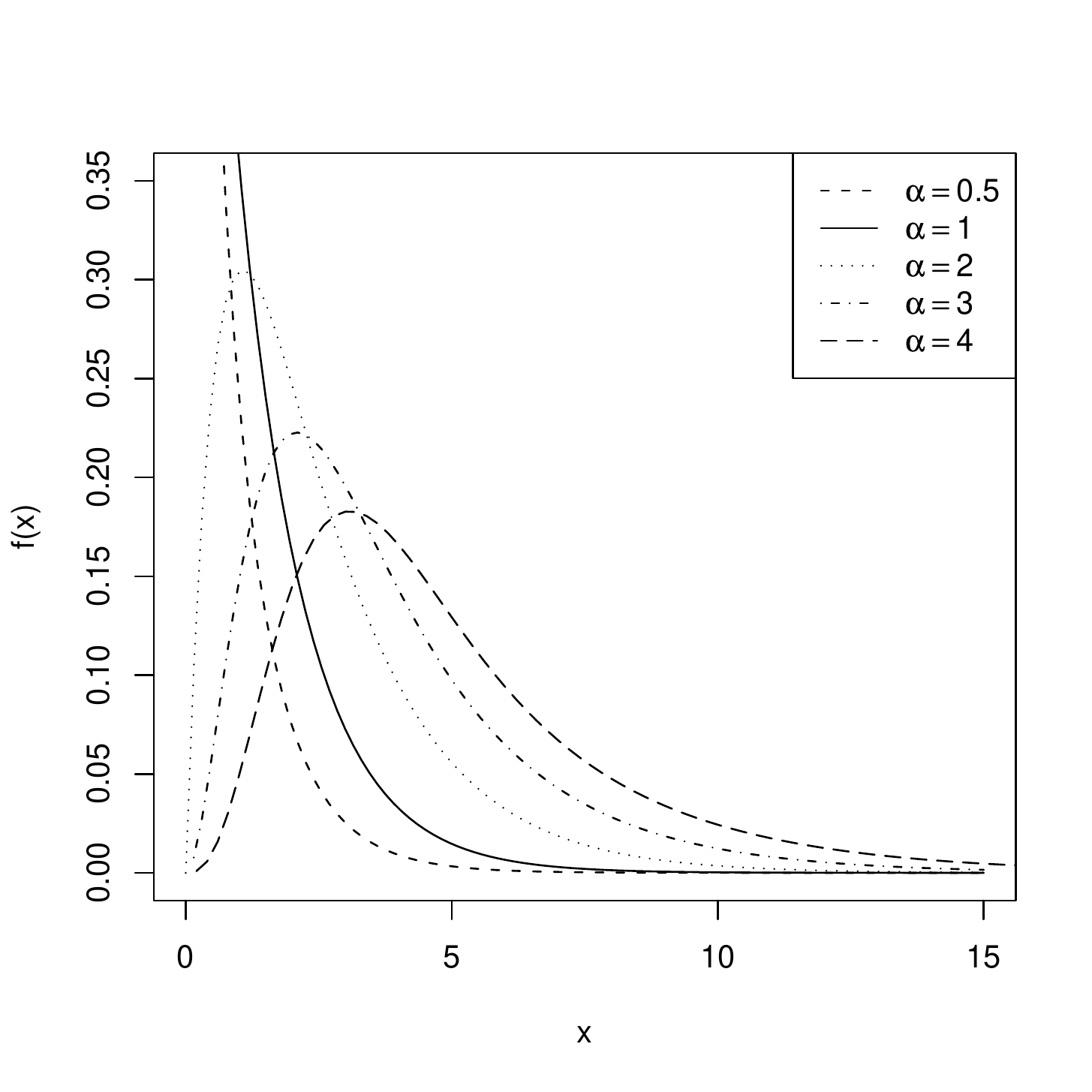}
	\caption{Plots of the marginal density of $X$ for $\beta=1$, $\alpha=0.5,1,2,3,4$, $p=0.2$ (left) and $p=0.8$ (right).}
	\label{marginaldensities}
\end{figure}

The marginal density of $X$ with respect to Lebesgue measure is an infinite mixture of gamma densities, which is given by 
\begin{eqnarray}\label{marginal}
f_X(x)=\sum_{n=1}^\infty P(N=n)g(x;n\alpha,\beta)=\frac{px^{-1}e^{-\beta x}}{1-p}\sum_{n=1}^\infty\frac{[(\beta x)^\alpha(1-p)]^n}{\Gamma(n\alpha)}, \quad x>0.
\end{eqnarray}
Therefore, the BGG distribution has infinite mixture of gamma and geometric marginals. Some alternative expressions for the marginal density of $X$ can be obtained. For example, for $\alpha=1$, we obtain the exponential density. Further, with help from Wolfram\footnote{http://www.wolframalpha.com/}, for $\alpha=1/2, 2, 3,4$, we have that
$$f_X(x)=p\beta^{1/2} x^{-1/2}e^{-\beta x}\{a(x)e^{a(x)^2}(1+\mbox{erf}(a(x)))+\pi^{-1/2}\},$$

$$f_X(x)=\frac{p\beta e^{-\beta x}}{\sqrt{1-p}}\sinh(\beta x\sqrt{1-p}),$$

$$f_X(x)=\frac{px^{-1} e^{-\beta x}}{3(1-p)}a(x)^{1/3}e^{-a(x)^{1/3}/2}\{e^{3a(x)^{1/3}/2}-2\sin(1/6(3\sqrt{3}a(x)^{1/3}+\pi))\},$$
and
$$f_X(x)=\frac{px^{-1} e^{-\beta x}}{2(1-p)}a(x)^{1/4}\{\sinh(a(x)^{1/4})-\sin(a(x)^{1/4})\},$$
respectively, where $a(x)\equiv a(x;\beta,\alpha,p)=(1-p)(\beta x)^\alpha$ and $\mbox{erf}(x)=2\pi^{-1/2}\int_0^xe^{-t^2/2}dt$ is the error function.
Figure \ref{marginaldensities} shows some plots of the marginal density of $X$ for $\beta=1$, $p=0.2,0.8$ and some values of $\alpha$.

We now obtain some conditional distributions which may be useful in goodness-of-fit analyses when the BGG distribution is assumed to model real data (see Section 4). Let $m\leq n$ be positive integers and $x>0$. The conditional cdf of $(X,N)$ given $N\leq n$ is 
$$P(X\leq x,N\leq m|N\leq n)=\frac{p}{1-(1-p)^n}\sum_{j=1}^m(1-p)^{j-1}\frac{\Gamma_{\beta x}(j\alpha)}{\Gamma(j\alpha)}.$$
We have that $P(X\leq x| N\leq n)$ is given by the right side of the above expression with $n$ replacing $m$.

For $0<x\leq y$ and $n\in\mathbb{N}$, the conditional cdf of $(X,N)$ given $X\leq y$ is
$$P(X\leq x,N\leq n|X\leq y)=\frac{\sum_{j=1}^n(1-p)^{j-1}\Gamma_{\beta x}(j\alpha)/\Gamma(j\alpha)}{\sum_{j=1}^\infty (1-p)^{j-1}\Gamma_{\beta y}(j\alpha)/\Gamma(j\alpha)}.$$

The conditional probability $P(N\leq n|X\leq y)$ is given by the right side of the above expression with $y$ replacing $x$.

From (\ref{density}) and (\ref{marginal}), we obtain that the conditional probability mass function of $N$ given $X=x$ is
\begin{eqnarray*}\label{NgivenX}
P(N=n|X=x)=\frac{[(1-p)(\beta x)^{\alpha}]^n/\Gamma(\alpha n)}{\sum_{j=1}^\infty[(1-p)(\beta x)^{\alpha}]^j/\Gamma(j\alpha)},
\end{eqnarray*}
for $n\in\mathbb{N}$. If $\alpha$ is known, the above probability mass function belongs to the one-parameter power series class of distributions; for instance, see Noack (1950). In this case, the parameter would be $(1-p)(\beta x)^\alpha$. For $\alpha=1$, we obtain the Poisson distribution truncated at zero with parameter $\beta x(1-p)$, which agrees with formula (7) from Kozuboswki and Panorska (2005). For the choice $\alpha=2$, we have that 
$$P(N=n|X=x)=\frac{(1-p)^{n-1/2}(\beta x)^{2n-1}}{(2n-1)!\sinh(\beta x\sqrt{1-p})},$$
where $n\in\mathbb{N}$.

\subsection{Moments}

\begin{figure}[h]
	\centering
		\includegraphics[width=0.40\textwidth]{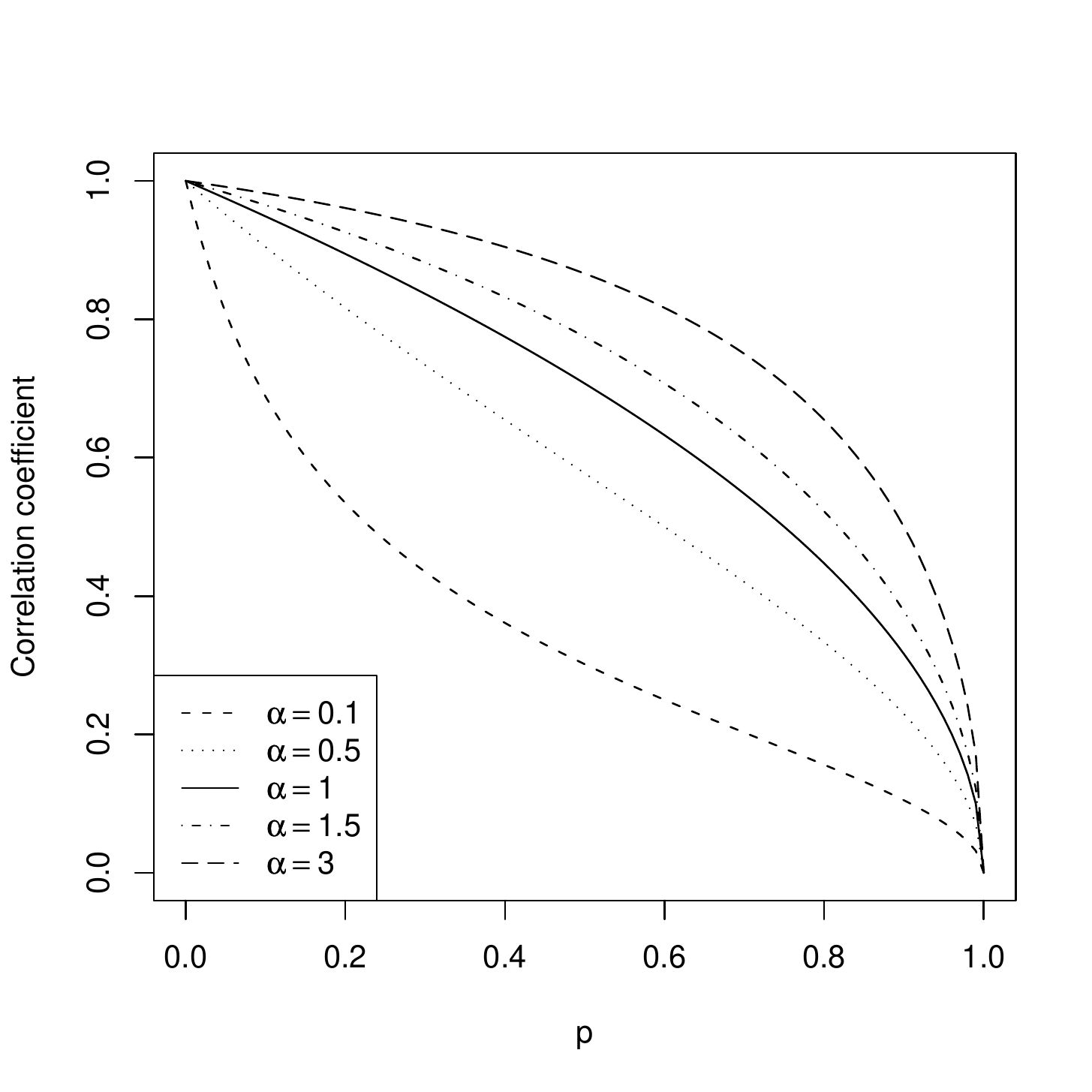}
	\caption{Plots of the correlation coefficient of the BGG law as a function of $p$ for $\alpha=0.1,0.5,1,1.5,3$.}
	\label{coefcorr}
\end{figure}

The moment generation function (mgf) of the BGG law is
\begin{eqnarray*}
\varphi(t,s)=E\left(e^{tX+sN}\right)=E\left[e^{sN}E\left(e^{tX}|N\right)\right]=E\left\{\left[e^s\left(\frac{\beta}{\beta-t}\right)^\alpha\right]^N\right\}, \quad t<\beta,\, s\in\mathbb{R},
\end{eqnarray*}
and then
\begin{eqnarray}\label{mgf}
\varphi(t,s)=\frac{pe^s\beta^\alpha}{(\beta-t)^\alpha-e^s\beta^\alpha(1-p)},
\end{eqnarray}
for $t<\beta\{1-[(1-p)e^s]^{1/\alpha}\}$.
The characteristic function may be obtained in a similar way and is given by 
\begin{eqnarray}\label{cf}
\Phi(t,s)=\frac{pe^{is}\beta^\alpha}{(\beta-it)^\alpha-e^{is}\beta^\alpha(1-p)},
\end{eqnarray}
for $t,s\in\mathbb{R}$. With this, the product and marginal moments can be obtained by computing $E(X^mN^k)=\partial^{m+k}\varphi(t,s)\partial t^m\partial s^k|_{t,s=0}$ or $E(X^mN^k)=(-i)^{m+k}\partial^{m+k}\Phi(t,s)\partial t^m\partial s^k|_{t,s=0}$. Hence, we obtain the following expression for the product moments of the random vector $(X,N)$:
\begin{eqnarray}\label{pm}
E(X^mN^k)=\frac{p\Gamma(m)}{\beta^m}\sum_{n=0}^\infty\frac{n^k(1-p)^{n-1}}{B(\alpha n,m)},
\end{eqnarray}
where $B(a,b)=\Gamma(a)\Gamma(b)/\Gamma(a+b)$, for $a,b>0$, is the beta function. 
In particular, we obtain that $E(X)=\alpha(p\beta)^{-1}$, $E(N)=p^{-1}$ and the covariance matrix $\Sigma$
of $(X,N)$ is given by
\begin{eqnarray}\label{cov}
\Sigma=
\left(\begin{array}{ll}
\frac{(1-p)\alpha^2}{p^2\beta^2}+\frac{\alpha}{\beta^2p} & \frac{(1-p)\alpha}{\beta p^2}\\
 \frac{(1-p)\alpha}{\beta p^2} & \frac{1-p}{p^2}\\
\end{array}\right).
\end{eqnarray}
The correlation coefficient $\rho$ between $X$ and $N$ is $\rho=\sqrt{(1-p)/(1-p+p/\alpha)}$. Let $\rho^*=\sqrt{1-p}$, that is, the correlation coefficient of a bivariate random vector following BEG law. For $\alpha\leq1$, we have $\rho\leq\rho^*$, and for $\alpha>1$, it follows that $\rho>\rho^*$. Figure \ref{coefcorr} shows some plots of the correlation coefficient of the BGG law as a function of $p$ for some values of $\alpha$.

From (\ref{mgf}), we find that the marginal mgf of $X$ is given by
$$\varphi(t)=\frac{p\beta^\alpha}{(\beta-t)^\alpha-\beta^\alpha(1-p)},$$
for $t<\beta\{1-(1-p)^{1/\alpha}\}$. The following expression for the $r$th moment of $X$ can be obtained from above formula or (\ref{pm}):
$$E(X^r)=\frac{p\Gamma(r)}{\beta^r}\sum_{n=0}^\infty\frac{(1-p)^{n-1}}{B(\alpha n,r)}.$$
We notice that the above expression is valid for any real $r>0$.

\subsection{Infinitely divisibility, geometric stability and representations}

We now show that BGG law is infinitely divisible, just as BEG law is. Based on Kozubowski and Panorska (2005), we define the bivariate random vector $$(R,v)=\left(\sum_{i=1}^{1+nT}G_{i},\frac{1}{n}+T\right),$$
where the $G_{i}$'s are iid random variables following $\Gamma(\alpha/n,\beta)$ distribution and independent of the random variable $T$, which follows negative binomial $\mbox{NB}(r,p)$ distribution with the probability mass function
\begin{eqnarray}\label{nbpf}
P(T=k)=\frac{\Gamma(k+r)}{k!\Gamma(r)}p^r(1-p)^k, \quad k\in\mathbb{N}\cup\{0\},
\end{eqnarray}
where $r=1/n$.
The moment generation function of $(R,v)$ is given by
\begin{eqnarray*}\label{idmgf}
E\left(e^{tR+sv}\right)&=&E\left[e^{s/n+sT}E\left(e^{t\sum_{i=1}^{1+nT}G_i}\big|T\right)\right]\\
&=&e^{s/n}\left(\frac{\beta}{\beta-t}\right)^{\alpha/n}E\left\{\left[e^s\left(\frac{\beta}{\beta-t}\right)^\alpha\right]^T\right\}\\
&=&\left\{\frac{pe^s\beta^\alpha}{(\beta-t)^\alpha-e^s\beta^\alpha(1-p)}\right\}^r,
\end{eqnarray*}
which is valid for $t<\beta\{1-[(1-p)e^s]^{1/\alpha}\}$ and $s\in\mathbb{R}$.
In a similar way, we obtain that the characteristic function is given by 
\begin{eqnarray}\label{idcf}
E(e^{itR+isv})=\left\{\frac{pe^{is}\beta^\alpha}{(\beta-it)^\alpha-e^{is}\beta^\alpha(1-p)}\right\}^r, 
\end{eqnarray}
for $t,s\in\mathbb{R}$.
With this, we have that $E(e^{itR+isv})=\Phi(t,s)^{1/n}$, where $\Phi(t,s)$ is the characteristic function of the BGG law given in (\ref{cf}). In words, we have that BGG distribution is infinitely divisible.

The exponential, geometric and BEG distributions are closed under geometric summation. We now show that our distribution also enjoys this geometric stability property. Let $\{(X_i,N_i)\}_{i=1}^\infty$ be iid random vectors following $\mbox{BGG}(\beta,\alpha,p)$ distribution independent of $M$, where $M\sim\mbox{Geom}(q)$, with $0<q<1$. By using (\ref{mgf}) and the probability generation function of the geometric distribution, one may easily check that $$\sum_{i=1}^M(X_i,N_i)\sim\mbox{BGG}(\beta,\alpha,pq).$$
From the above result, we find another stochastic representation of the BGG law, which generalizes proposition (4.2) from Kozubowski and Panorska (2005):
$$(X,N)\stackrel{d}{=}\sum_{i=1}^M(X_i,N_i),$$
where $\{(X_i,N_i)\}_{i=1}^\infty\stackrel{iid}{\sim}\mbox{BGG}(\beta,\alpha,p/q)$, with $0<p<q<1$, and $M$ is defined as before. In what follows, another representation of the BGG law is provided, by showing that it is a convolution of a bivariate distribution (with gamma and degenerate at 1 marginals) and a compound Poisson distribution. Let $\{Z_i\}_{i=1}^\infty$ be a sequence of iid random variables following logarithmic distribution with probability mass function $P(Z_i=k)=(1-p)^k(\lambda k)^{-1}$, for $k\in\mathbb{N}$, where $\lambda=-\log p$. Define the random variable $Q\sim\mbox{Poisson}(\lambda)$, independent of the $Z_i$'s. Given the sequence $\{Z_i\}_{i=1}^\infty$, let $G_i\sim\Gamma(\alpha Z_i,\beta)$, for $i\in\mathbb{N}$, be a sequence of independent random variables and let $G\sim\Gamma(\alpha,\beta)$ be independent of all previously defined variables. Then, we have that
\begin{eqnarray}\label{cp}
(X,N)\stackrel{d}{=}(G,1)+\sum_{i=1}^{Q}(G_i,Z_i).
\end{eqnarray}
Taking $\alpha=1$ in (\ref{cp}), we obtain the proposition 4.3 from Kozubowski and Panorska (2005). To show that the above representation holds, we use the probability generation functions $E\left(t^Q\right)=e^{\lambda(t-1)}$ (for $t\in\mathbb{R}$) and $E\left(t^{Z_i}\right)=\log(1-(1-p)t)/\log p$ (for $t<(1-p)^{-1}$). With this, it follows that
\begin{eqnarray}\label{cp1}
E\left(e^{t(G+\sum_{i=1}^QG_i)+s(1+\sum_{i=1}^QZ_i)}\right)&=&e^s\left(\frac{\beta}{\beta-t}\right)^\alpha E\left\{\left[E\left(e^{tG_1+sZ_1}\right)\right]^Q\right\}\nonumber\\
&=&e^s\left(\frac{\beta}{\beta-t}\right)^\alpha e^{\lambda\left\{E\left(e^{tG_1+sZ_1}\right)-1\right\}},
\end{eqnarray}
for $t<\beta$. Furthermore, for $t<\beta\{1-[(1-p)e^s]^{1/\alpha}\}$, we have that 
\begin{eqnarray*}\label{cp2}
E\left(e^{tG_1+sZ_1}\right)=E\left\{\left[\frac{e^s\beta^\alpha}{(\beta-t)^\alpha}\right]^{Z_1}\right\}=\frac{\log\{1-(1-p)e^s\beta^\alpha/(\beta-t)^\alpha\}}{\log p}.
\end{eqnarray*}
By using the above result in (\ref{cp1}), we obtain the representation (\ref{cp}).

\section{Estimation and inference}

Let $(X_1,N_1)$, \ldots, $(X_n,N_n)$ be a random sample from $\mbox{BGG}(\beta,\alpha,p)$ distribution and $\theta=(\beta,\alpha,p)^\top$ be the parameter vector. The log-likelihood function $\ell=\ell(\theta)$ is given by
\begin{eqnarray}\label{loglik}
\ell&\propto& n\alpha\log\beta\,\bar{N}_n+n\log p-n\beta\bar{X}_n+n\log(1-p)(\bar{N}_n-1)\nonumber\\
&&+\sum_{i=1}^n\left\{\alpha N_i\log X_i-\log\Gamma(\alpha N_i)\right\},
\end{eqnarray}
where $\bar{X}_n=\sum_{i=1}^nX_i/n$ and $\bar{N}_n=\sum_{i=1}^nN_i/n$. The associated score function $U(\theta)=(\partial\ell/\partial\beta,\partial\ell/\partial\alpha,\partial\ell/\partial p)^\top$ to log-likelihood function (\ref{loglik}) comes
\begin{eqnarray*}
\frac{\partial\ell}{\partial\beta}=n\left(\frac{\alpha\bar{N}_n}{\beta}-\bar{X}_n\right),
\quad \frac{\partial\ell}{\partial\alpha}=n\bar{N}_n\log\beta +\sum_{i=1}^n\left\{N_i\log X_i-N_i\Psi(\alpha N_i)\right\}
\end{eqnarray*}
and 
\begin{eqnarray}\label{scorep}
\frac{\partial\ell}{\partial p}=\frac{n}{p}-\frac{n(\bar{N}_n-1)}{1-p},
\end{eqnarray}
where $\Psi(x)=d\log\Gamma(x)/dx$. By solving the nonlinear system of equations $U(\Theta)=0$, it follows that the maximum likelihood estimators (MLEs) of the parameters are obtained by
\begin{eqnarray}\label{mles}
\widehat\beta=\widehat\alpha\frac{\bar{N}_n}{\bar{X}_n},\quad \widehat{p}=\frac{1}{\bar{N}_n}\quad\mbox{and}\quad \sum_{i=1}^nN_i\Psi(\widehat\alpha N_i)-n\bar{N}_n\log\left(\frac{\widehat\alpha \bar{N}_n}{\bar{X}_n}\right)=\sum_{i=1}^nN_i\log X_i.
\end{eqnarray}
Since MLE of $\alpha$ may not be found in closed-form, nonlinear optimization algorithms such as a Newton algorithm or a quasi-Newton
algorithm are needed.

We are now interested in constructing confidence intervals for the parameters. For this, the Fisher's information matrix is required. The information matrix $J(\theta)$ is
\begin{eqnarray}\label{inform}
J(\theta)=
\left(\begin{array}{lll}
\kappa_{\beta\beta} & \kappa_{\beta\alpha} & 0\\
\kappa_{\beta\alpha} & \kappa_{\alpha\alpha} & 0\\
0 & 0 & \kappa_{pp} \\
\end{array}\right),
\end{eqnarray}
with $$\kappa_{\beta\beta}=\frac{\alpha}{\beta^2p},\quad \kappa_{\beta\alpha}=-\frac{1}{\beta p},\quad \kappa_{\alpha\alpha}=p\sum_{j=1}^\infty j^2(1-p)^{j-1}\Psi'(j\alpha) \quad \mbox{and} \quad \kappa_{pp}=\frac{1}{p^2(1-p)}.$$
where $\Psi'(x)=d\Psi(x)/dx$. 

Standard large sample theory gives us that $\sqrt{n}(\widehat\theta-\theta)\stackrel{d}{\rightarrow}N_3\left(0,J^{-1}(\theta)\right)$ as $n\rightarrow\infty$, where $J^{-1}(\theta)$ is the inverse matrix of $J(\theta)$ defined in (\ref{inform}). 

The asymptotic multivariate normal distribution of $\sqrt{n}(\widehat\theta-\theta)$ can be used to construct
approximate confidence intervals and confidence regions for the parameters. Further, we can compute the maximum values of the unrestricted and restricted log-likelihoods to construct the likelihood ratio (LR) statistic for testing some sub-models of the BGG distribution. For example,
we may use the LR statistic for testing the hypotheses $H_0 \mbox{:} \,\,\alpha=1$ versus $H_1 \mbox{:} \,\,\alpha\neq1$, which corresponds to test BEG distribution versus BGG distribution.

\subsection{A reparametrization}

We here propose a reparametrization of the bivariate gamma-geometric distribution and show its advantages over the previous one. Consider the reparametrization $\mu=\alpha/\beta$ and $\alpha$ and $p$ as before. Define now the parameter vector $\theta^*=(\mu,\alpha,p)^\top$. Hence, the density (\ref{density}) now becomes 
\begin{eqnarray*}
f^*_{X,N}(x,n)=\frac{(\alpha/\mu)^{n\alpha}}{\Gamma(\alpha n)}x^{n\alpha-1}e^{-\alpha x/\mu}p(1-p)^{n-1}, \quad x>0,\,\, n\in\mathbb{N}.
\end{eqnarray*} 
We shall denote $(X,N)\sim\mbox{BGG}(\mu,\alpha,p)$. Therefore if $(X_1,N_1)$, \ldots, $(X_n,N_n)$ is a random sample from $\mbox{BGG}(\mu,\alpha,p)$
distribution, the log-likelihood function $\ell^*=\ell(\theta^*)$ is given by
\begin{eqnarray}\label{loglikr}
\ell^*&\propto& n\alpha\log\left(\frac{\alpha}{\mu}\right)\bar{N}_n+n\log p-n\frac{\alpha}{\mu}\bar{X}_n+n\log(1-p)(\bar{N}_n-1)\nonumber\\
&&+\sum_{i=1}^n\left\{\alpha N_i\log X_i-\log\Gamma(\alpha N_i)\right\}.
\end{eqnarray}
The score function associated to (\ref{loglikr}) is $U^*(\theta^*)=(\partial\ell^*/\partial\mu,\partial\ell^*/\partial\alpha,\partial\ell^*/\partial p)^\top$, where 
\begin{eqnarray*}
\frac{\partial\ell^*}{\partial\mu}=\frac{n\alpha}{\mu}\left(\frac{\bar{X}_n}{\mu}-\bar{N}_n\right), \quad \frac{\partial\ell^*}{\partial\alpha}=n\bar{N}_n\log\left(\frac{\alpha}{\mu}\right)+\sum_{i=1}^nN_i\{\log X_i-\Psi(\alpha N_i)\}
\end{eqnarray*}
and $\partial\ell^*/\partial p$ is given by (\ref{scorep}). The MLE of $p$ is given (as before) in (\ref{mles}), and the MLEs of $\mu$ and $\alpha$ are obtained by $$\widehat\mu=\frac{\bar{X}_n}{\bar{N}_n}\quad\mbox{and}\quad \sum_{i=1}^nN_i\Psi(\widehat\alpha N_i)-n\bar{N}_n\log\left(\widehat\alpha\frac{\bar{N}_n}{\bar{X}_n}\right)=\sum_{i=1}^nN_i\log X_i.$$
As before nonlinear optimization algorithms are needed to find MLE of $\alpha$. Under this reparametrization, Fisher's information matrix $J^*(\theta^*)$ becomes
\begin{eqnarray*}
J^*(\theta^*)=
\left(\begin{array}{lll}
\kappa^*_{\mu\mu} & 0 & 0\\
0 & \kappa^*_{\alpha\alpha} & 0\\
0 & 0 & \kappa^*_{pp} \\
\end{array}\right),
\end{eqnarray*}
with $$\kappa^*_{\mu\mu}=\frac{\alpha}{\mu^2p},\quad \kappa^*_{\alpha\alpha}=p\sum_{j=1}^\infty j^2(1-p)^{j-1}\Psi'(j\alpha)-\frac{1}{\alpha p}\quad \mbox{and} \quad \kappa^*_{pp}=\kappa_{pp}.$$
The asymptotic distribution of $\sqrt{n}(\widehat\theta^*-\theta^*)$ is trivariate normal with null mean and covariance matrix $J^{*\,-1}(\theta^*)=\mbox{diag}\{1/k^*_{\mu\mu}, 1/k^*_{\alpha\alpha},1/k_{pp}\}$. We see that under this reparametrization we have orthogonal parameters in the sense of Cox and Reid (1987); the information matrix is a diagonal matrix. With this, we obtain desirable properties such as asymptotic independence of the estimates of the parameters. The reader is referred to Cox and Reid (1987) for more details. 

\section{Application}

\begin{figure}[h!]
	\centering
		\includegraphics[width=0.33\textwidth]{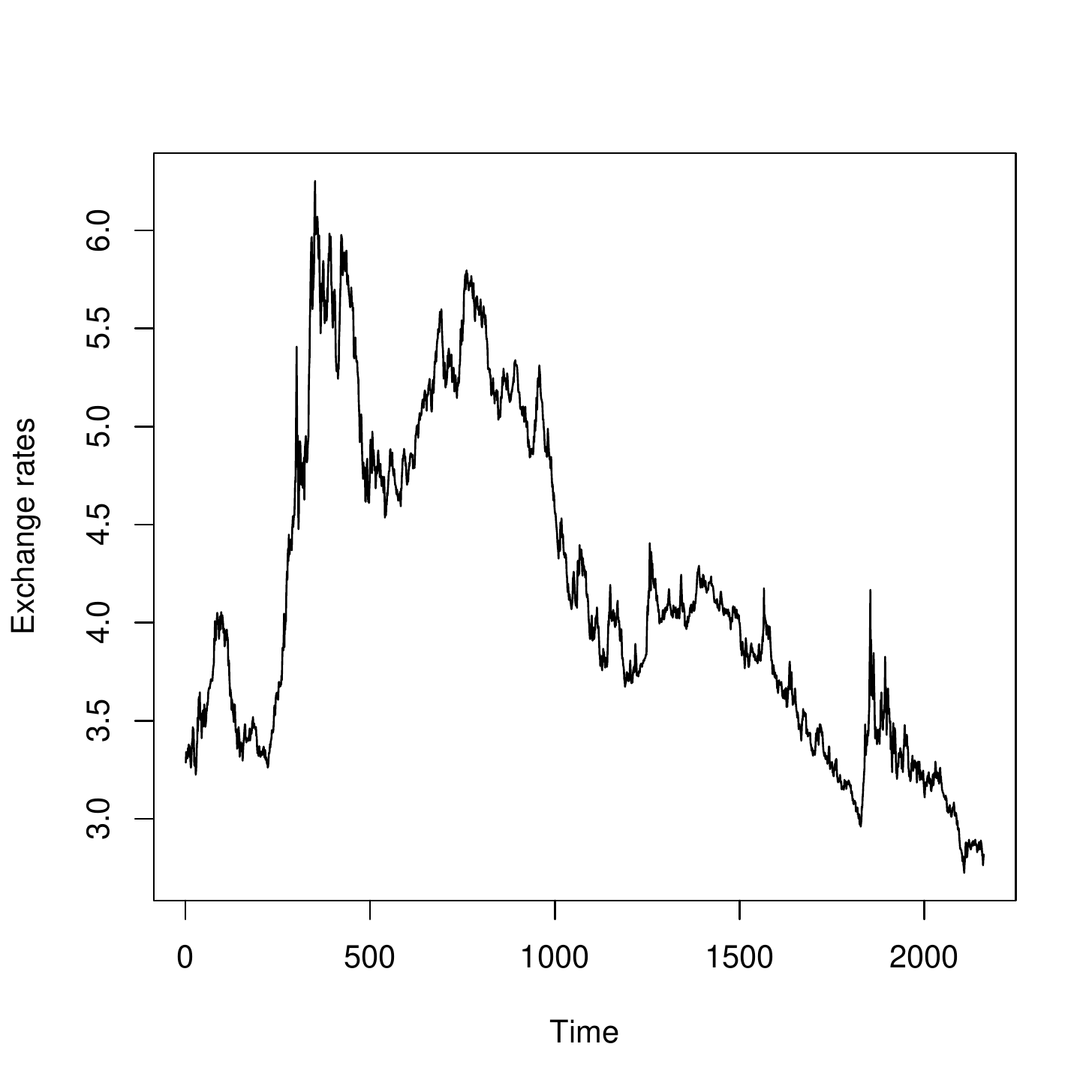}\includegraphics[width=0.33\textwidth]{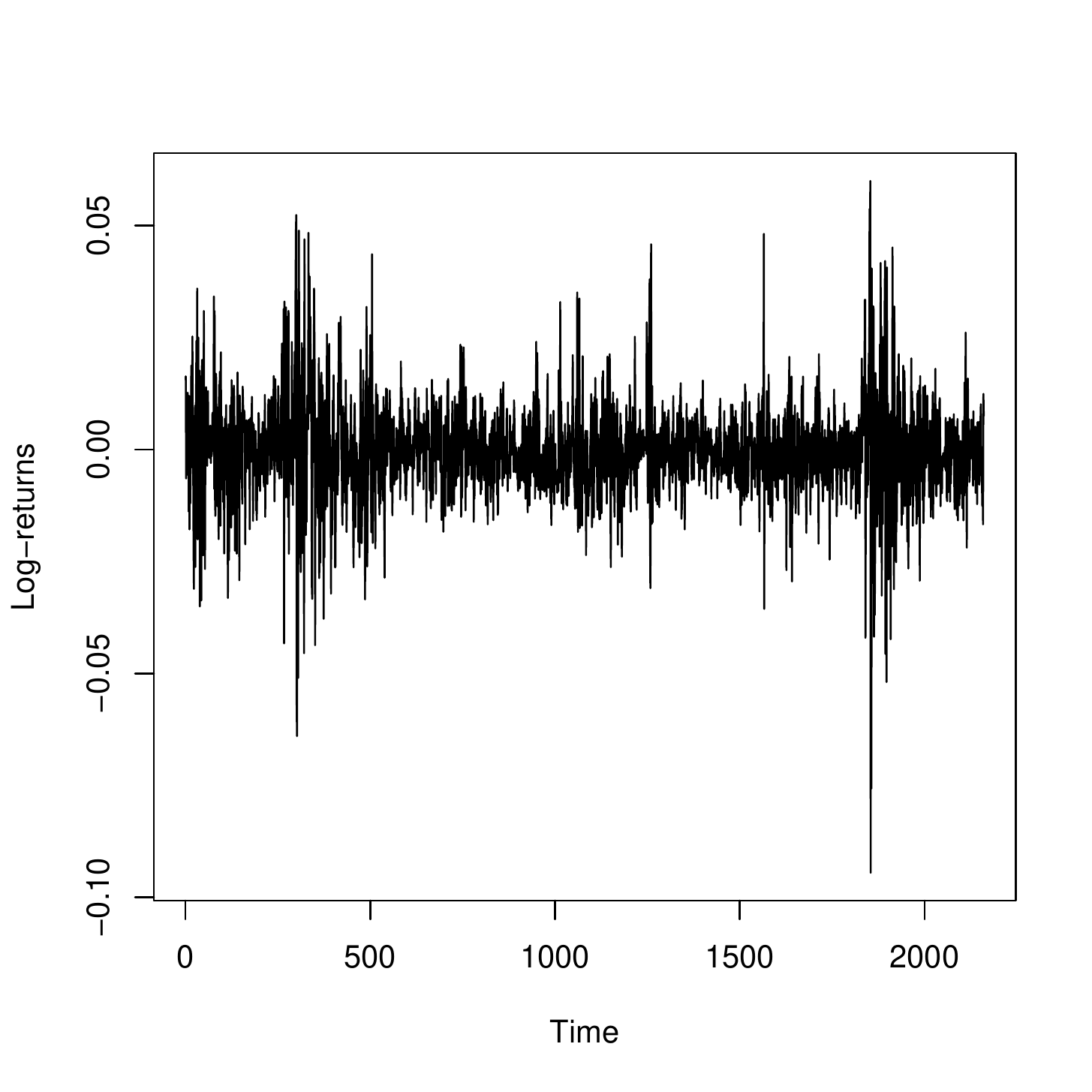}
	\caption{Graphics of the daily exchange rates and log-returns.}
	\label{log-returns}
\end{figure}

Here, we show the usefulness of the bivariate gamma-geometric law applied to a real data set. We consider daily exchange rates between Brazilian real and U.K. pounds, quoted in Brazilian real, covering May 22, 2001 to December 31, 2009. With this, we obtain the daily log-returns, that is, the logarithms of the rates between two consecutive exchange rates. Figure \ref{log-returns} illustrates the daily exchange rates and the log-returns.  

We will jointly model the magnitude and duration of the consecutive positive log-returns by using BGG law. We call attention that the duration of the consecutive positive log-returns is the same that the duration of the growth periods of the exchange rates. The data set consists of 549 pairs $(X_i,N_i)$, where $X_i$ and $N_i$ are the magnitude and duration as described before, for $i=1,\ldots,549$. We notice that this approach of looking jointly at the magnitude and duration of the consecutive positive log-returns was firstly proposed by Kozubowski and Panorska (2005) with the BEG model, which showed a good fit to another currencies considered. Suppose $\{(X_i,N_i)\}_{i=1}^{549}$ are iid random vectors following $\mbox{BGG}(\mu,\alpha,p)$ distribution. We work with the reparametrization proposed in the Subsection 3.1. 

Table \ref{summaryfit} presents a summary of the fit of our model, which contains maximum likelihood estimates of the parameters with their respective standard errors, and asymptotic confidence intervals at the 5\% significance level. Note that the confidence interval of $\alpha$ does not contain the value $1$. Then, for the Wald test, we reject the hypothesis that the data come from BEG distribution in favor of the BGG distribution, at the 5\% significance level. We also perform likelihood ratio (LR) test and obtain that the LR statistic is equal to $5.666$ with associated p-value $0.0173$. Therefore, for any usual significance level (for example 5\%), the likelihood ratio test rejects the hypothesis that the data come from BEG distribution in favor of the BGG distribution, so agreeing with Wald test's decision. The empirical and fitted correlation coefficients are equal to 0.6680 and 0.6775, respectively, therefore, we have a good agreement between them.

\begin{table}[h!]
\centering
\begin{tabular}{c|cccc}
\hline
Parameters & Estimate & Stand. error & Inf. bound & Sup. bound \\
\hline
$\mu$ & 0.0082 & 0.00026 & 0.0076 & 0.0087 \\ 
$\alpha$ &  0.8805 & 0.04788 & 0.7867 &  0.9743 \\
$p$ & 0.5093 & 0.01523 &  0.4794 & 0.5391 \\
\hline
\end{tabular}
\caption{Maximum likelihood estimates of the parameters, standard errors and bounds of the asymptotic confidence intervals at the 5\% significance level.}\label{summaryfit}
\end{table}

The BEG model was motived by an empirical observation that the magnitude of the consecutive positive log-returns followed the same type of distribution as the positive one-day log-returns (see Kozubowski and Panorska, 2005). Indeed, the marginal distribution of $X$ in the BEG model is also exponential (with mean $p^{-1}\mu$), just as the positive daily log-returns (with mean $\mu$). This stability of the returns was observed earlier by Kozubowski and Podg\'orski (2003), with the log-Laplace distribution. We notice that BGG distribution does not enjoy this stability property, since the marginal distribution of $X$ is an infinite mixture of gamma distributions. We now show that the data set considered here does not present this stability. 

\begin{figure}[h!]
	\centering
		\includegraphics[width=0.4\textwidth]{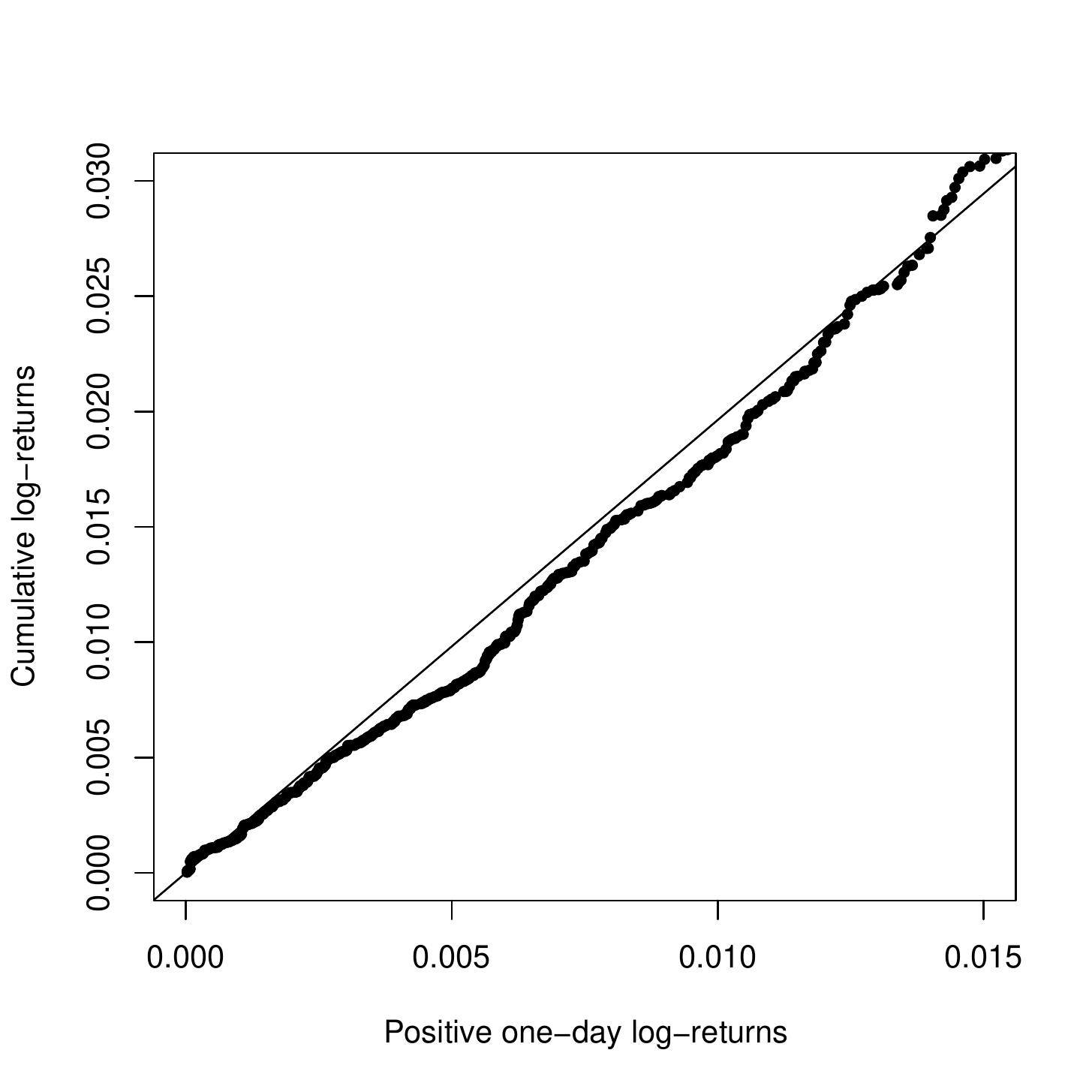}
	\caption{Empirical quantile-quantile plot between cumulative consecutive positive log-returns and positive one-day log-returns, with the straight line $y=1.9636 x$. The range $(x,y)\in(0,0.015)\times(0,0.030)$ covers 85\% of the data set.}
	\label{qqplot}
\end{figure}

Denote the $i$th positive one-day log-returns by $D_i$ and define $D_i^*=p^{-1}D_i$. If the data was generated from a $\mbox{BEG}(\mu,p)$ distribution, then an empirical quantile-quantile plot between the $X_i$'s ($y$-axis) and the $D_i$'s ($x$-axis) would be around the straight line $y=p^{-1}x$, for $x>0$. Figure \ref{qqplot} presents this plot and we observe that a considerable part of the points are below of the straight line $y=1.9636 x$ (we replace $p$ by its MLE $\widehat p=0.5093$). Therefore, the present data set seems to have been generated by a distribution that lacks the stability property discussed above. In order to confirm this, we test the hypothesis that the $X_i$'s and $D_i^*$'s have the same distribution. In the BEG model, both have exponential distribution with mean $\mu$. Since $\widehat p$ converges in probability to $p$ (as $n\rightarrow\infty$), we perform the test with $\widehat p$ replacing $p$. The Kolmogorov-Smirnov statistic and associated p-value are equal to 0.0603 and 0.0369, respectively. Therefore, using a significance level at 5\%, we reject the hypothesis that the $X_i$'s and $D_i^*$'s have the same distribution.

\begin{figure}[h!]
	\centering
		\includegraphics[width=0.33\textwidth]{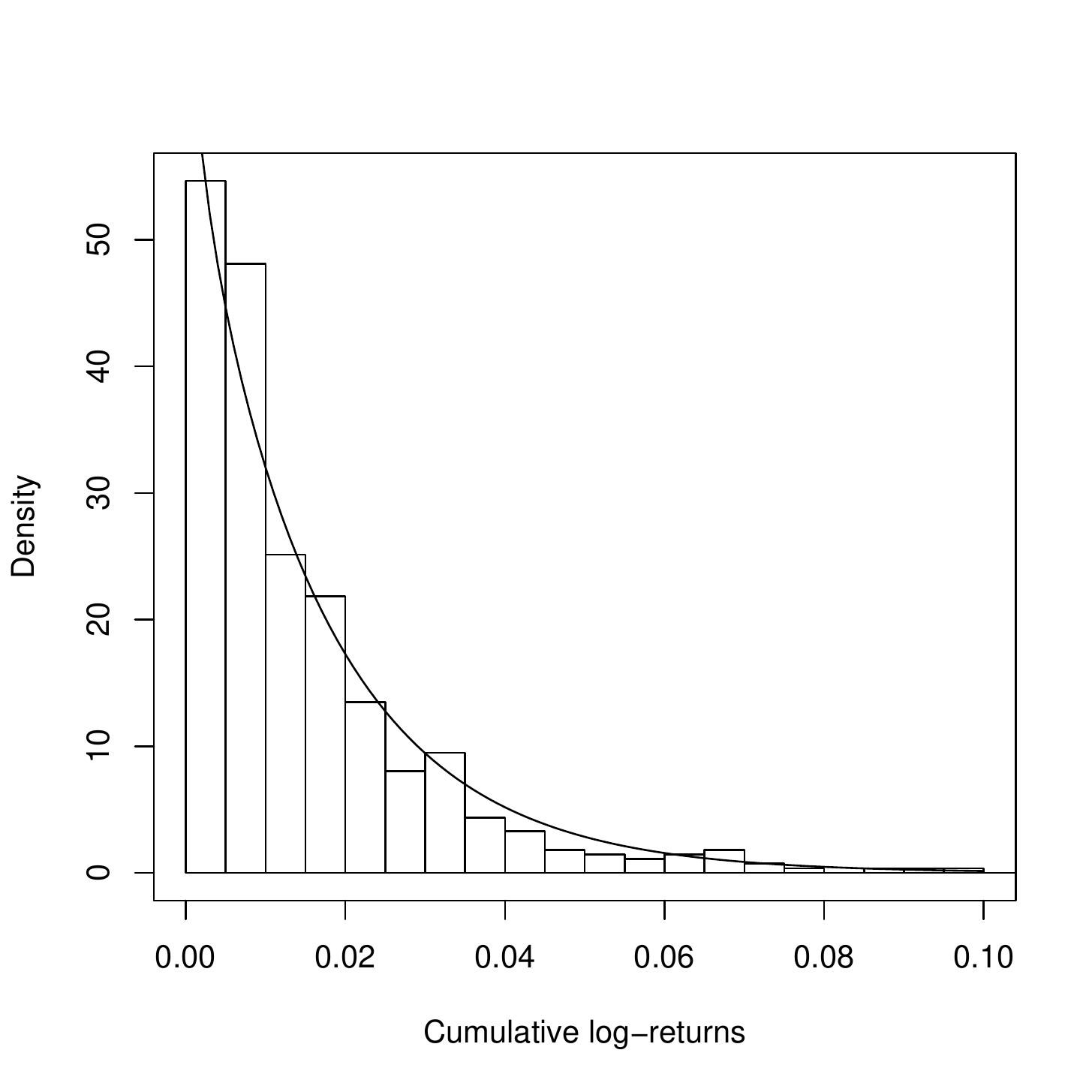}\includegraphics[width=0.33\textwidth]{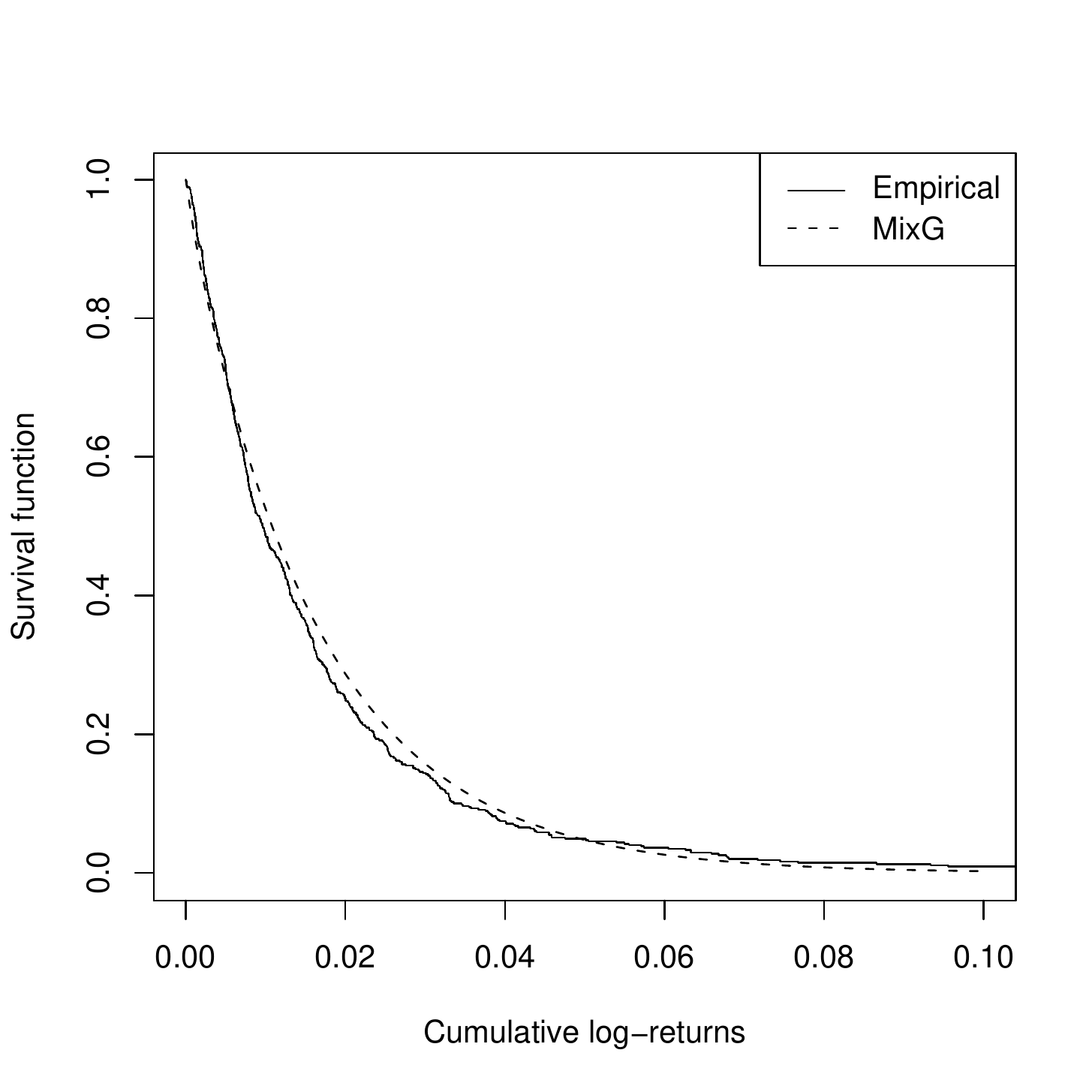}
	\caption{Plot on the left shows the fitted mixture of gamma densities (density of $X$) with the histogram of the data. Plot on the right presents the empirical and fitted theoretical (mixture of gamma) survival functions. }
	\label{fit_densities}
\end{figure}

Figure \ref{fit_densities} presents the fitted marginal density (mixture of gamma densities) of the cumulative log-returns with the histogram of the data and the empirical and fitted survival functions. These plots show a good fit of the mixture of gamma distributions to the data. This is confirmed by the Kolmogorov-Smirnov (KS) test, which we use to measure the goodness-of-fit of the mixture of gamma distributions to the data. The KS statistic and its p-value are equal to 0.0482 and 0.1557, respectively. Therefore, using any usual significance level, we accept the hypothesis that the mixture of gamma distributions is adequate to fit the cumulative log-returns. 

\begin{figure}[h!]
	\centering
		\includegraphics[width=0.33\textwidth]{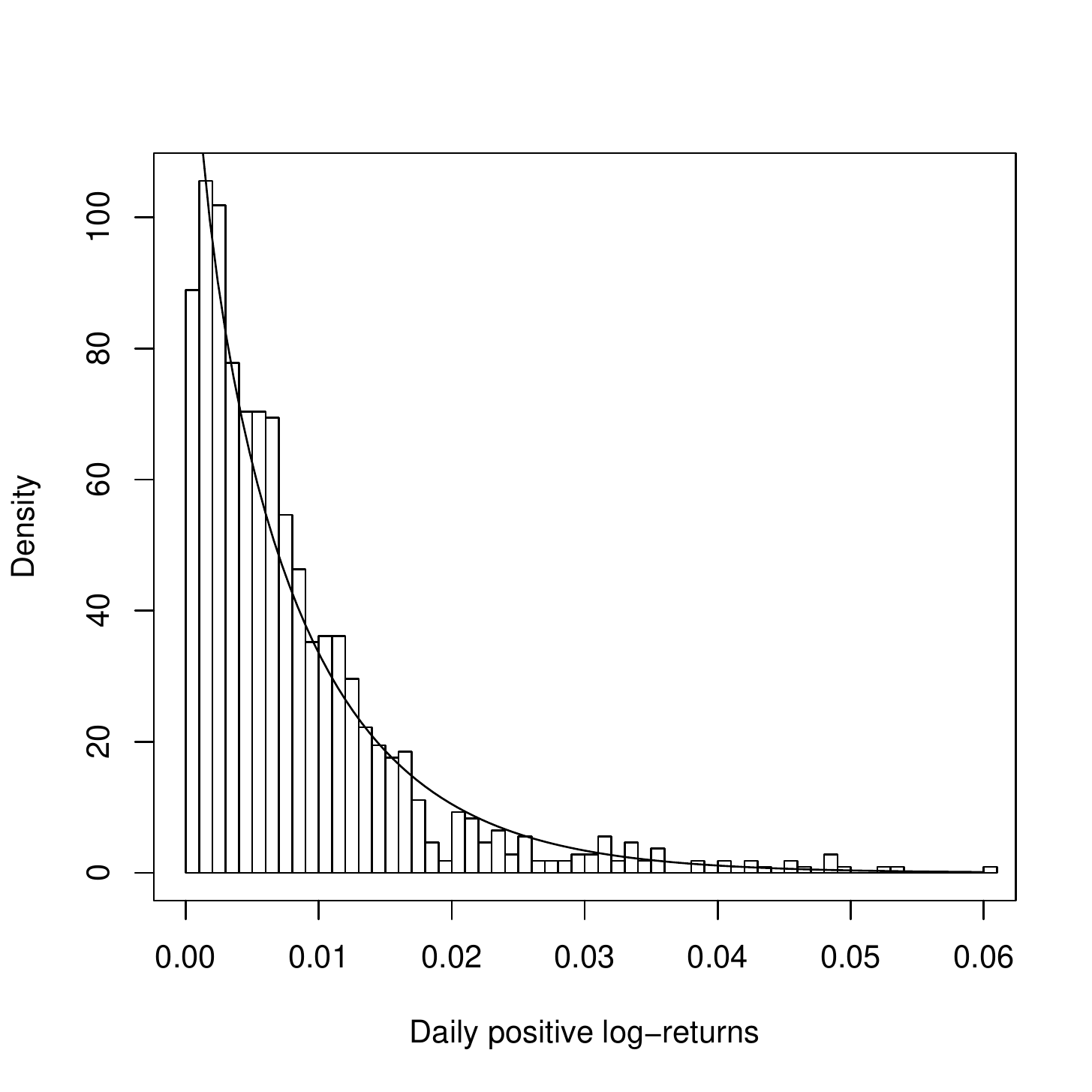}\includegraphics[width=0.33\textwidth]{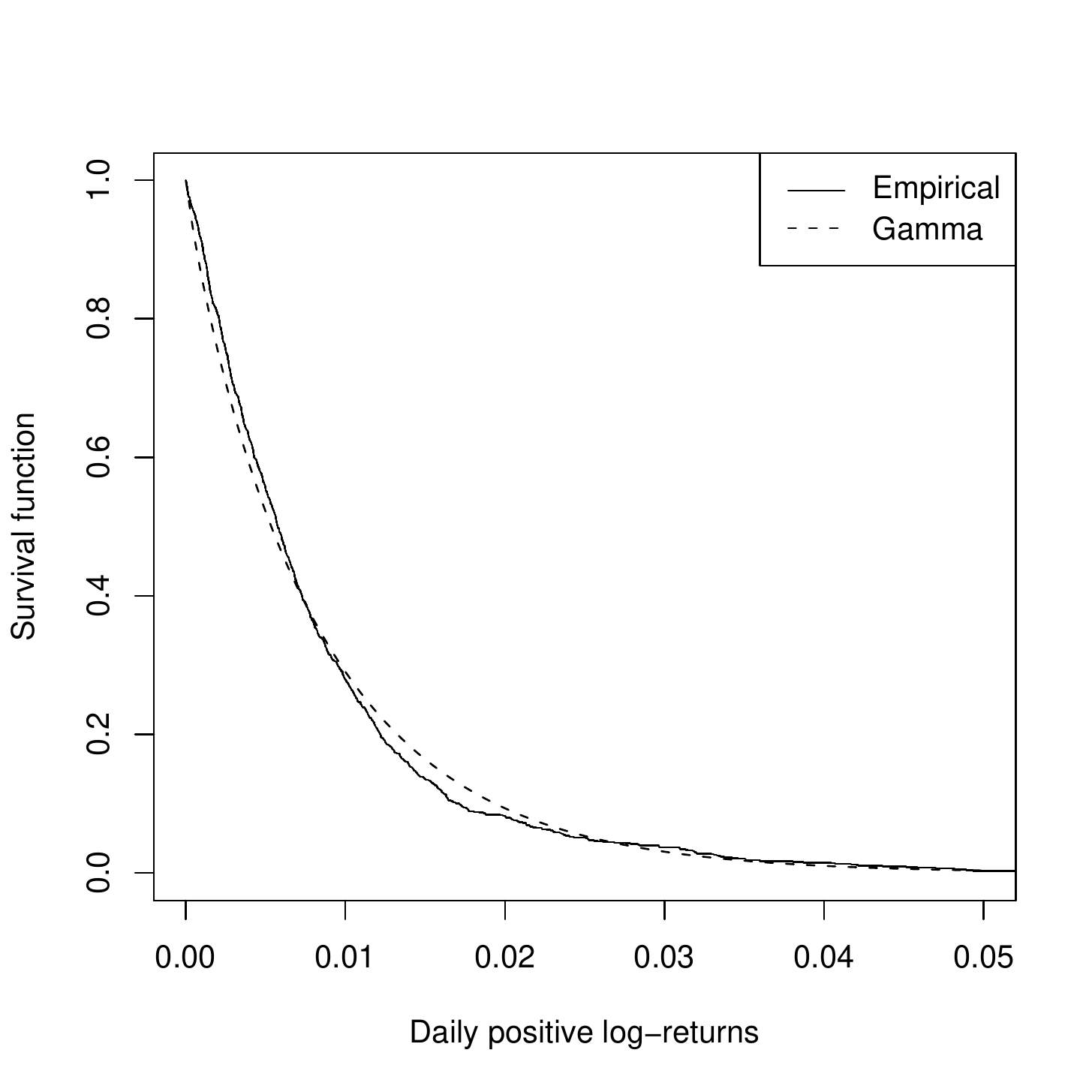}
	\caption{Picture on the left shows the histogram and fitted gamma density for the daily positive log-returns. Empirical survival and fitted gamma survival are shown in the picture on the right.}
	\label{dailypositiveplots}
\end{figure}

Plots of the histogram, fitted gamma density and empirical and fitted survival functions for the daily positive log-returns are presented in the Figure \ref{dailypositiveplots}. The good performance of the gamma distribution may be seen by these graphics. In the Table \ref{marginalgeom}
we show absolute frequency, relative frequency and fitted geometric model for the duration in days of the consecutive positive log-returns. From this, we observe that the geometric distribution fits well the data. This is confirmed by the Pearson's chi-squared (denoted by $\chi^2$) test, where our null hypothesis is that the duration follows geometric distribution. The $\chi^2$ statistic equals 42 (degrees of freedom equals 36) with associated p-value 0.2270, so we accept (using any usual significance level) that the growth period follows geometric distribution. We notice that geometric distribution has also worked quite well for modeling the duration of the growth periods of exchange rates as part of the BEG model in Kozubowski and Panorska (2005).

\begin{table}[h!]
\centering
\begin{tabular}{c|ccccccc}
\hline
$N\rightarrow$ & 1 & 2 & 3 & 4 & 5 & 6 & $\geq7$ \\
\hline
Absolute frequency &269 & 136 & 85 & 34 & 15 & 6 & 4 \\
Relative frequency& 0.48998& 0.24772 & 0.15483 & 0.06193 & 0.02732 & 0.01093 & 0.00728 \\
Fitted model & 0.50928 & 0.24991 & 0.12264  & 0.06018 & 0.02953 & 0.01449 & 0.01396\\
\hline
\end{tabular}
\caption{Absolute and relative frequencies and fitted marginal probability mass function of $N$ (duration in days of the growth periods).}\label{marginalgeom}
\end{table}

\begin{figure}[h!]
	\centering
		\includegraphics[width=0.33\textwidth]{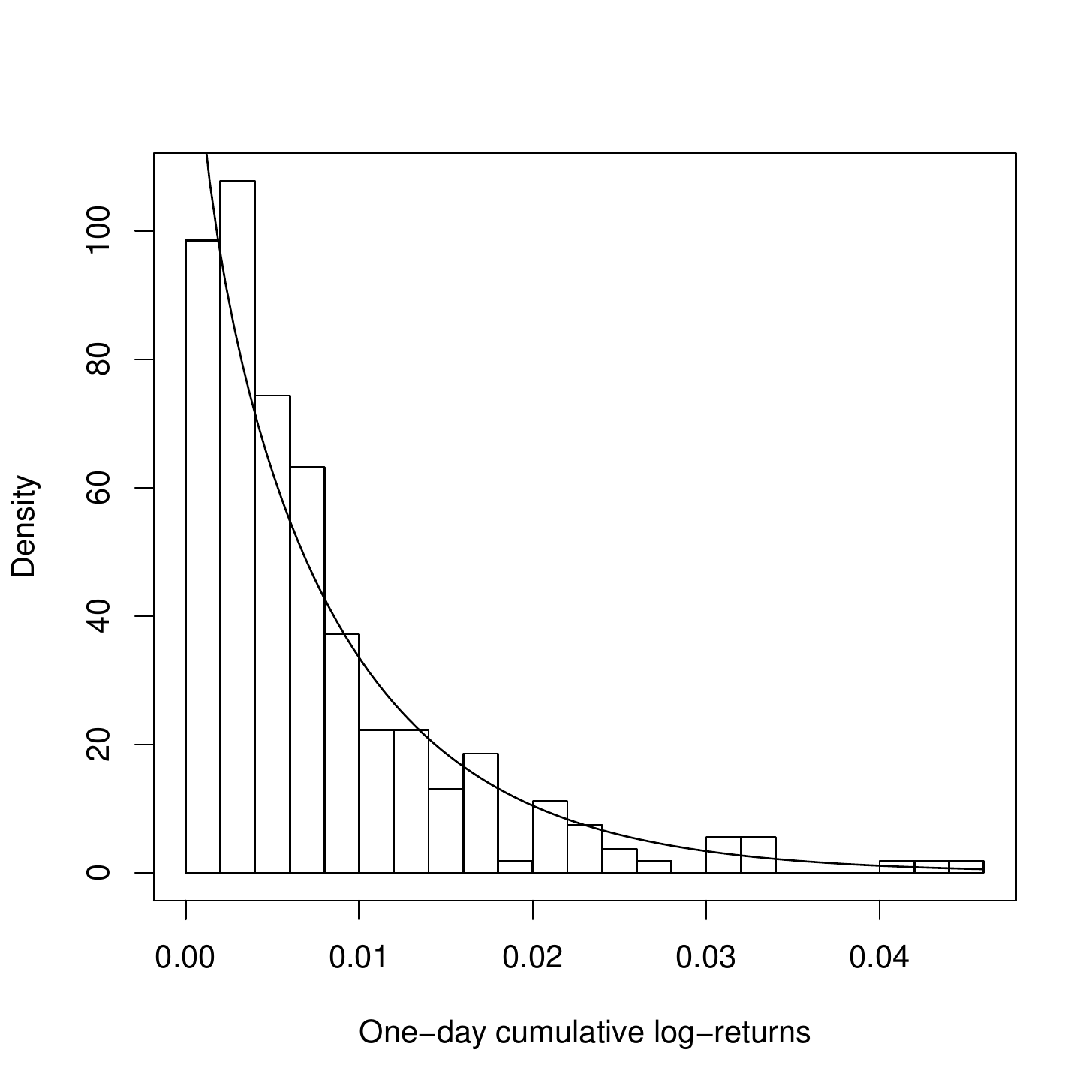}\includegraphics[width=0.33\textwidth]{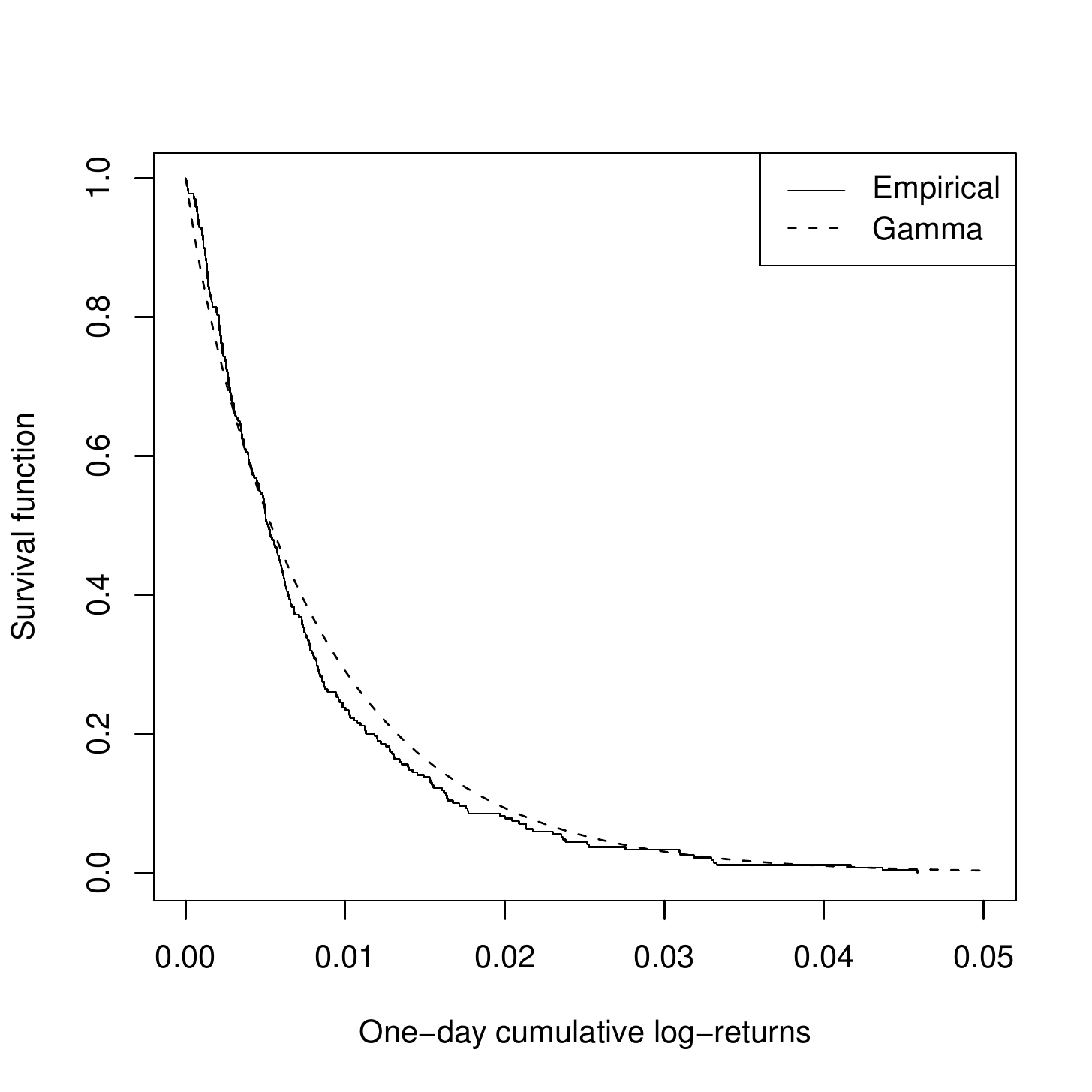}
		\includegraphics[width=0.33\textwidth]{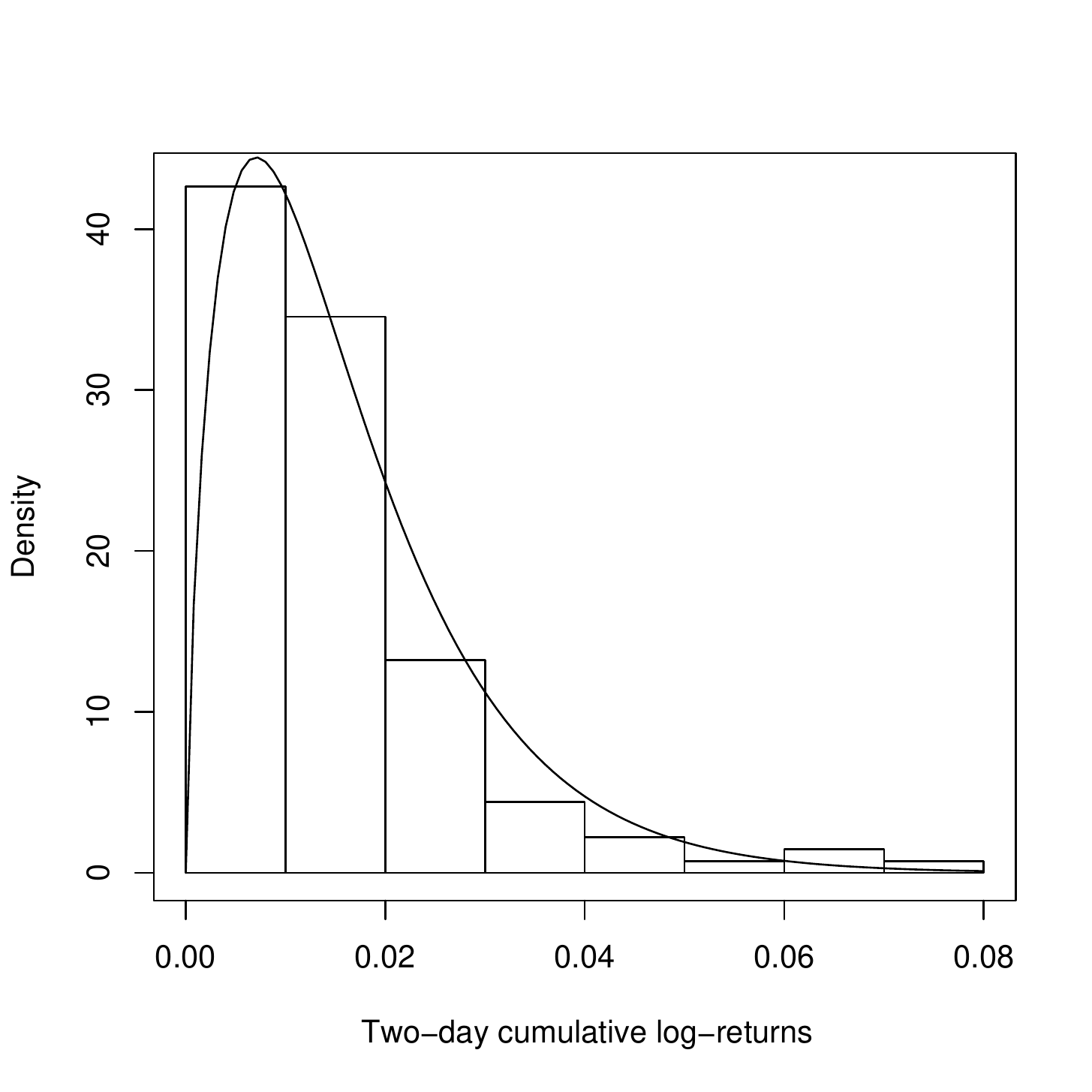}\includegraphics[width=0.33\textwidth]{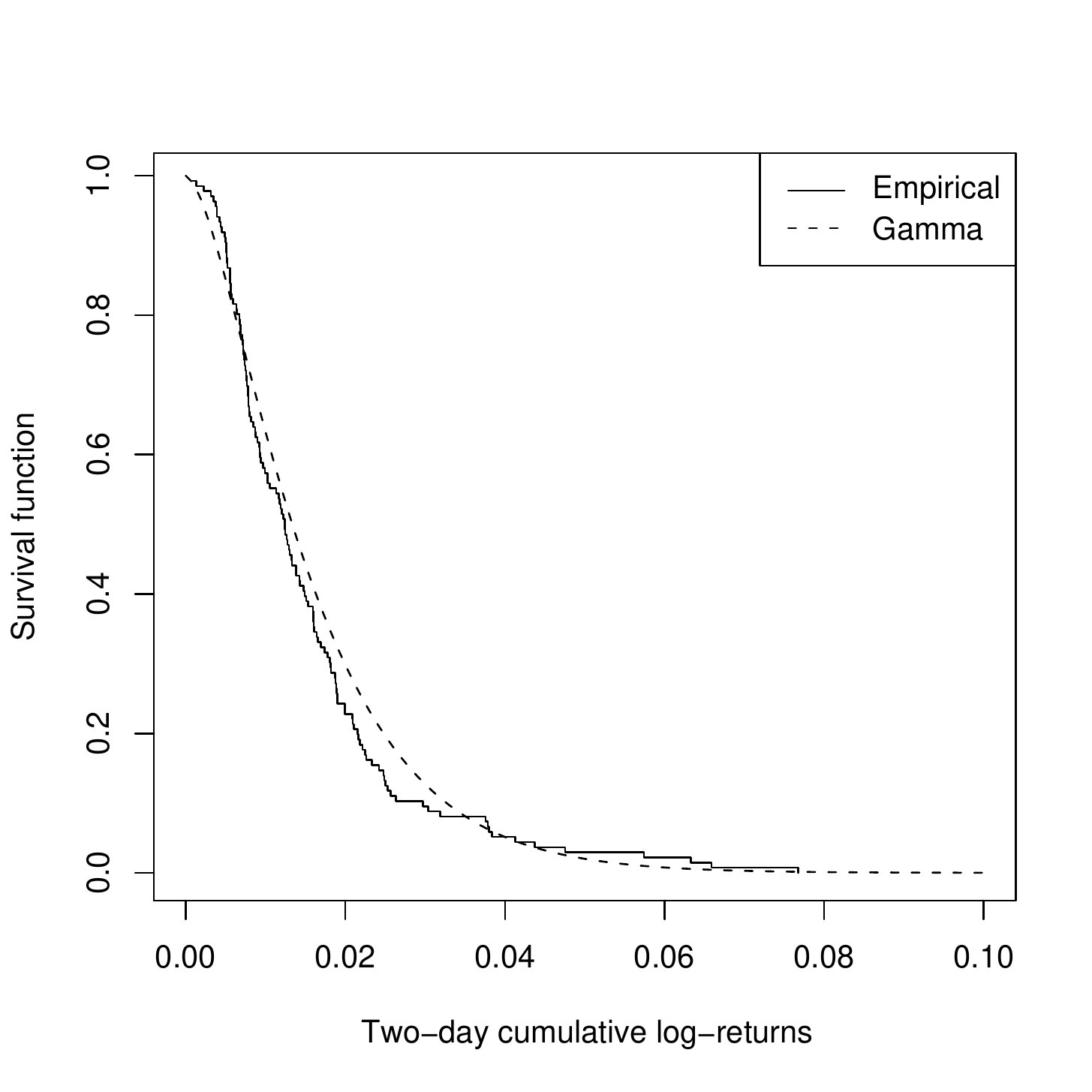}
		\includegraphics[width=0.33\textwidth]{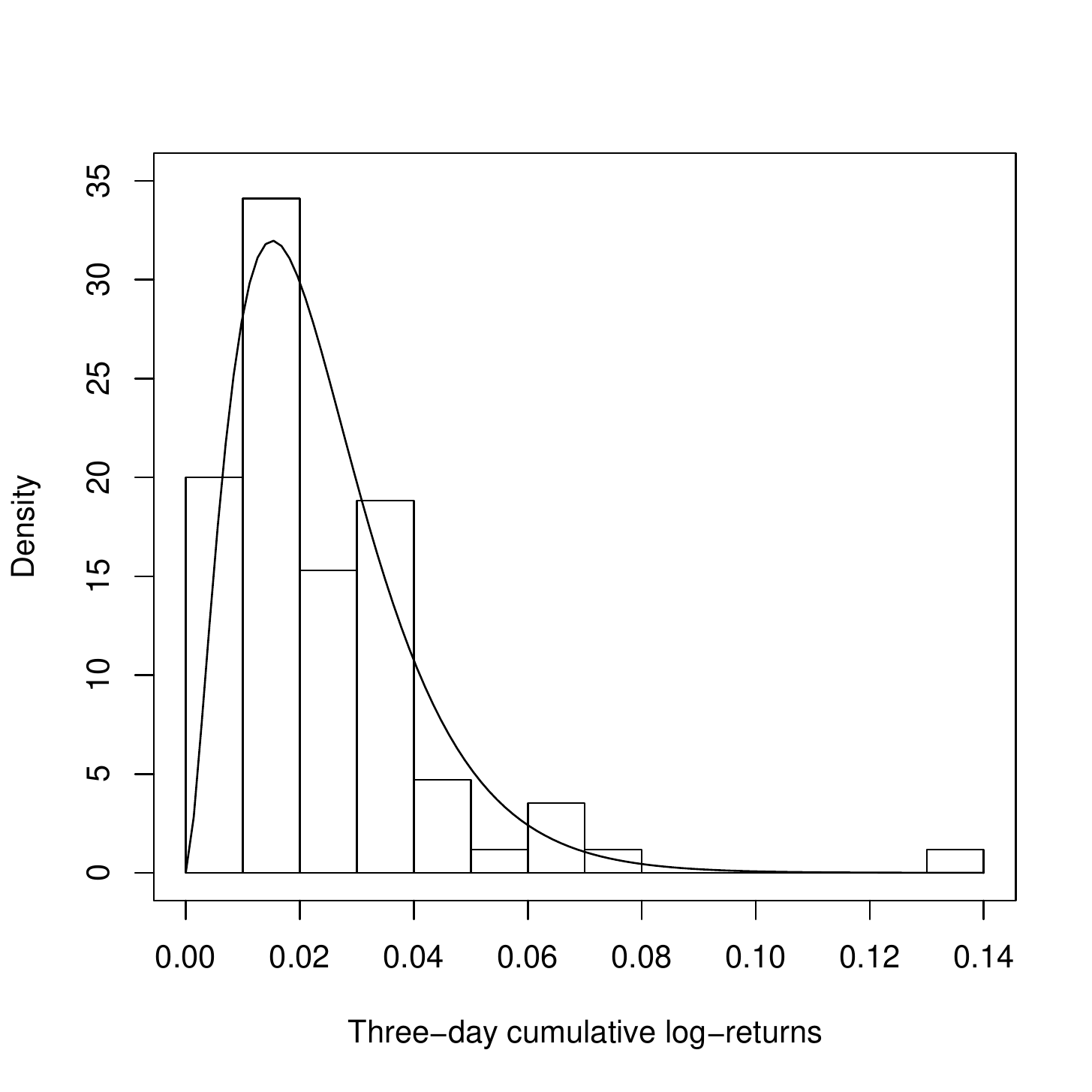}\includegraphics[width=0.33\textwidth]{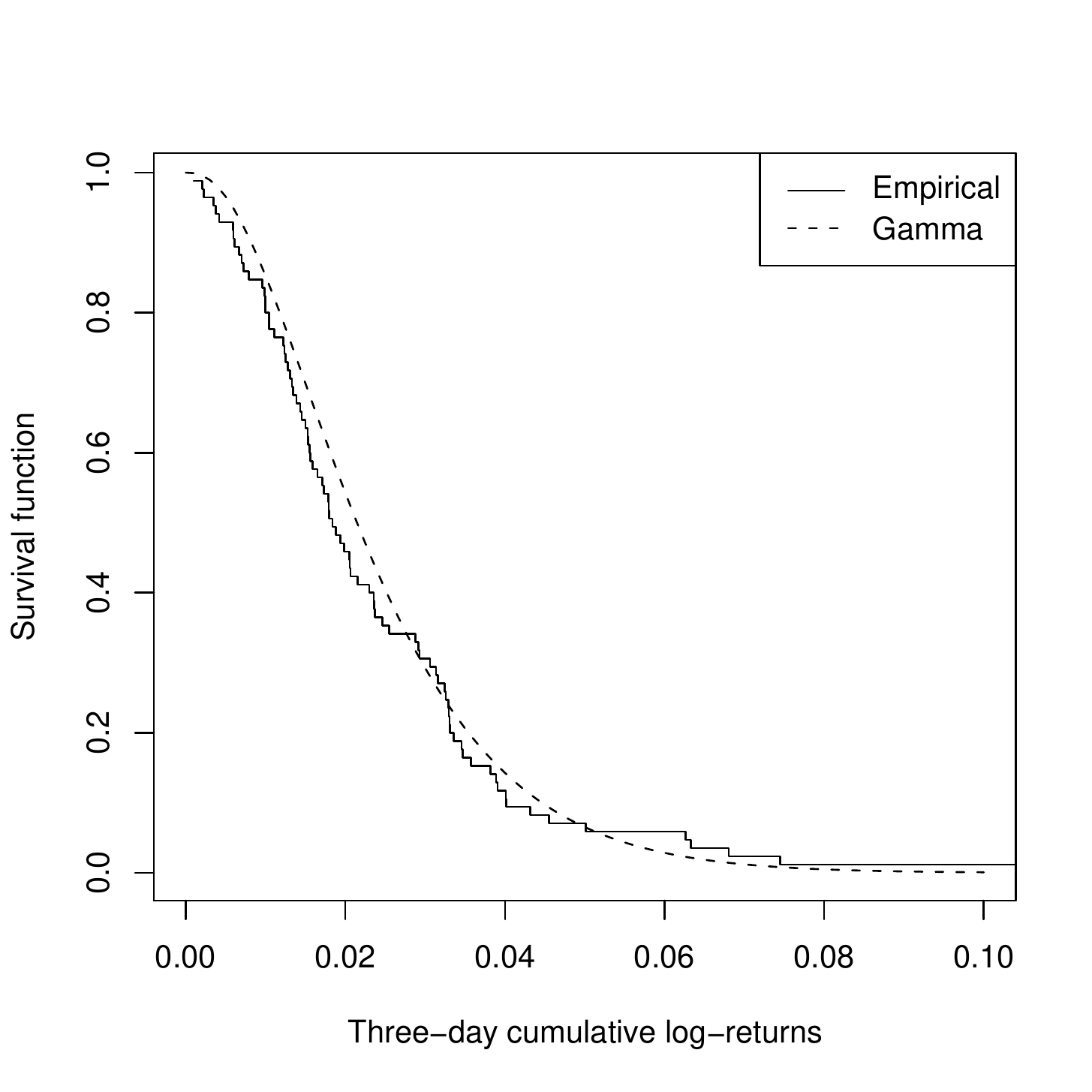}
		\caption{Plots of the fitted conditional density and survival functions of $X$ given $N=1$, $N=2$ and $N=3$. In the pictures of the density and survival functions, we also plot the histogram of the data and the empirical survival function, respectively.}
	\label{conditional_densities}
\end{figure}

\begin{figure}[h!]
	\centering
		\includegraphics[width=0.33\textwidth]{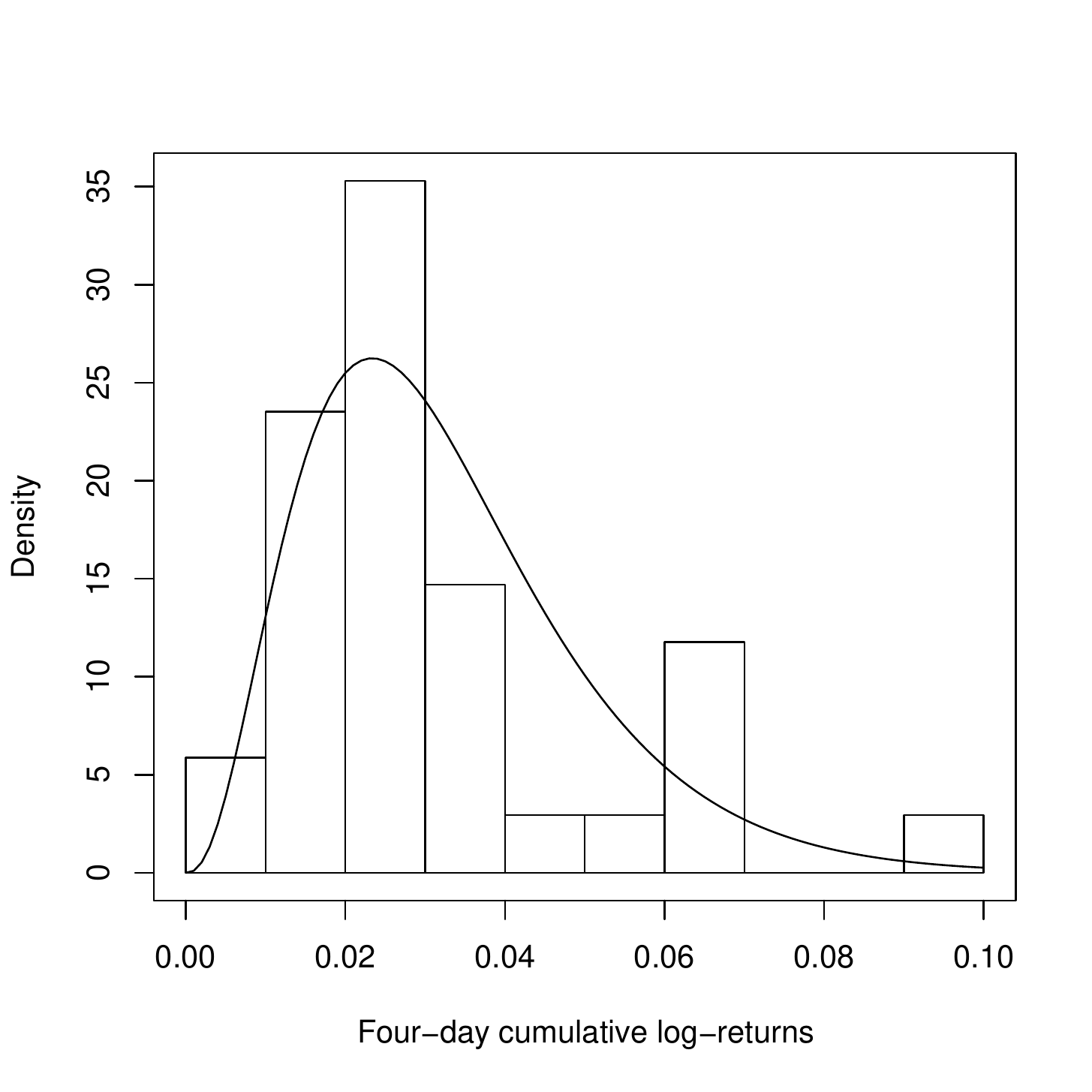}\includegraphics[width=0.33\textwidth]{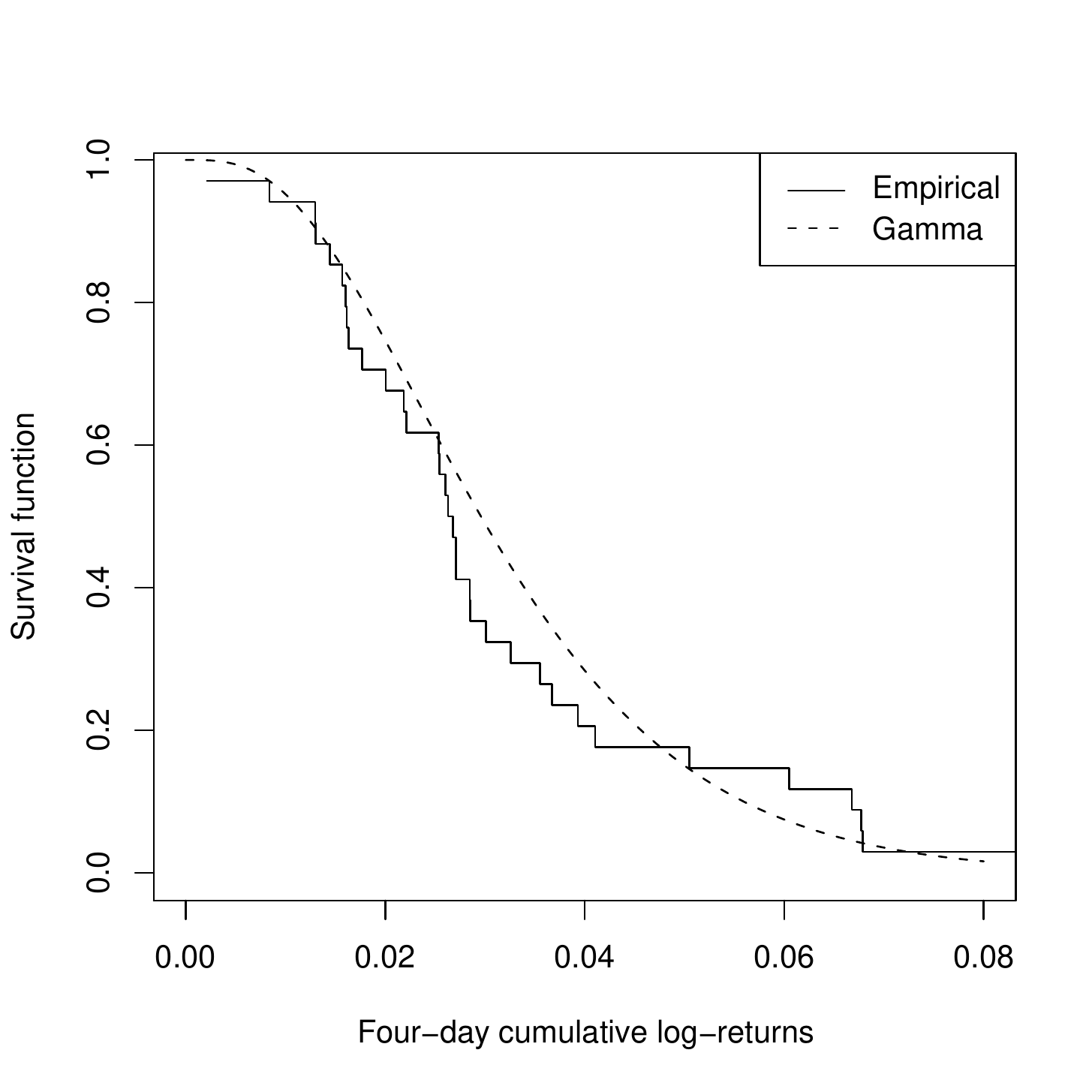}
		\includegraphics[width=0.33\textwidth]{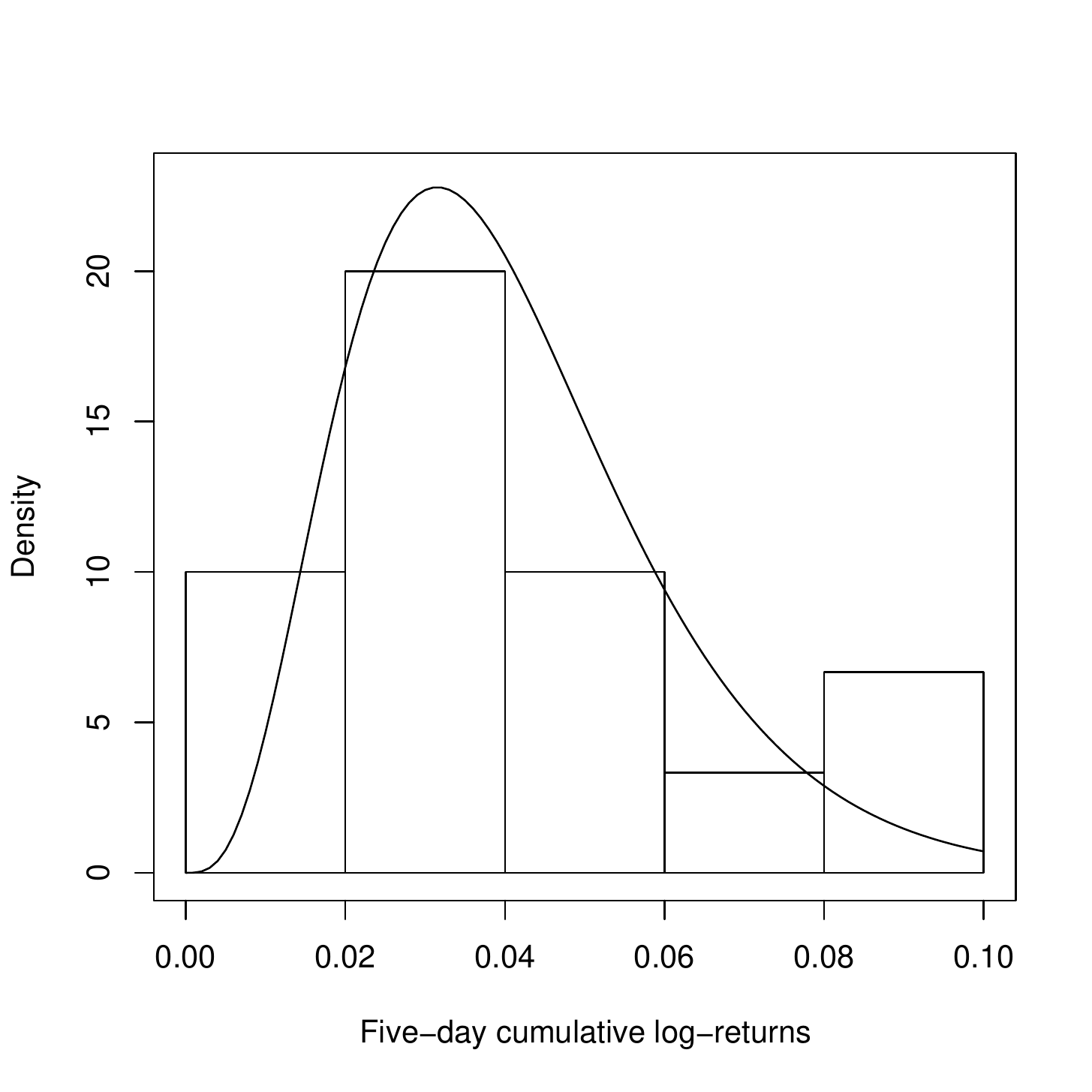}\includegraphics[width=0.33\textwidth]{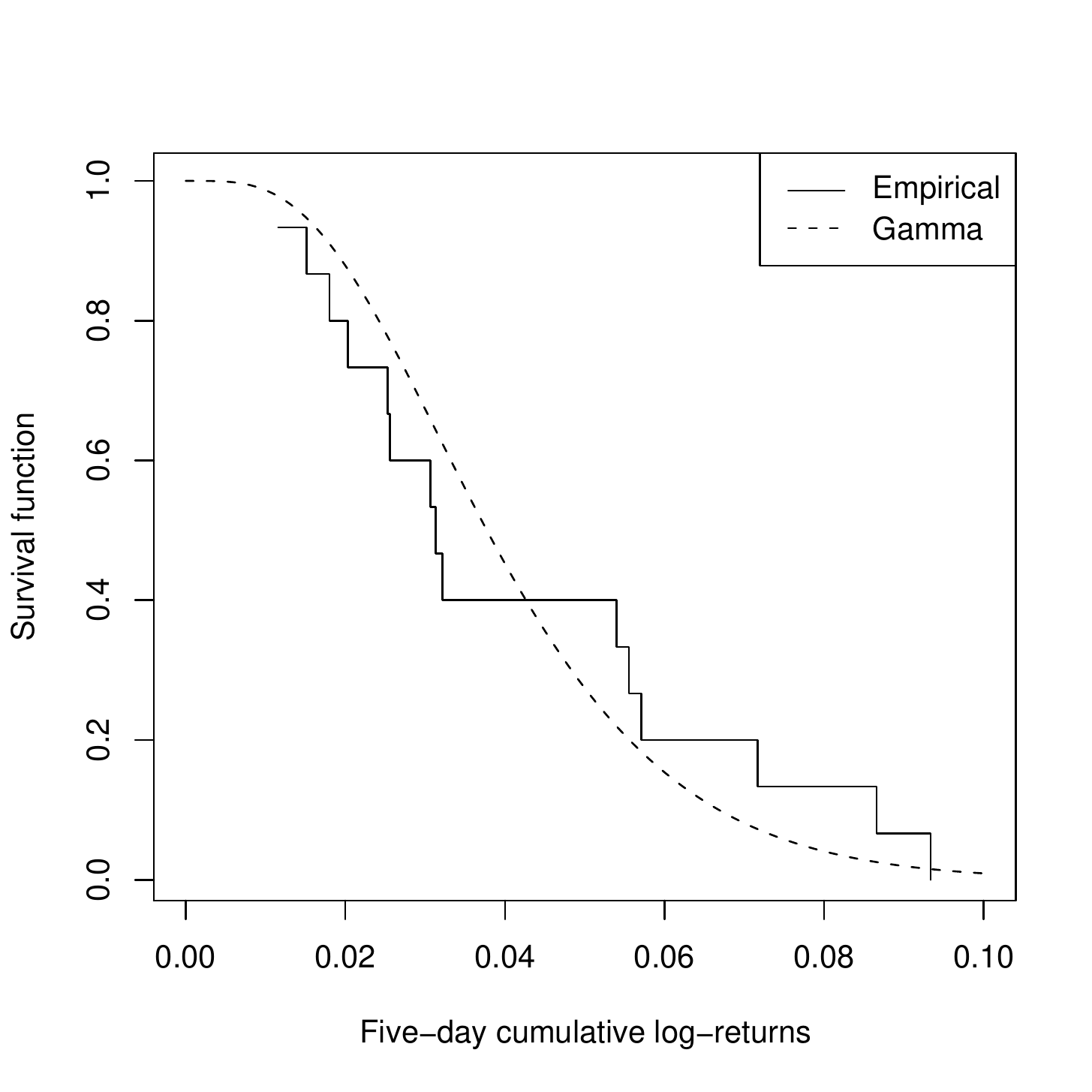}
	\caption{Plots of the fitted conditional density and survival functions of $X$ given $N=4$ and $N=5$. In the pictures of the density and survival functions, we also plot the histogram of the data and the empirical survival function, respectively.}
	\label{conditional_densities2}
\end{figure}

So far our analysis has showed that the bivariate gamma-geometric distribution and its marginals provided a suitable fit to the data. We end our analysis verifying if the conditional distributions of the cumulative log-returns given the duration also provide good fits to the data. As mentioned before, the conditional distribution of $X$ given $N=n$ is $\Gamma(n\alpha,\alpha/\mu)$. Figure \ref{conditional_densities} shows plots of the fitted density and fitted survival function of the conditional distributions of $X$ given $N=1,2,3$. The histograms of the data and the empirical survival functions are also displayed. The corresponding graphics for the conditional distributions of $X$ given $N=4,5$ are displayed in the Figure \ref{conditional_densities2}. 
These graphics show a good performance of the gamma distribution to fit cumulative log-returns given the growth period (in days). We also use the Kolmogorov-Smirnov test to verify the goodness-of-fit these conditional distributions. In the table \ref{conditionalX} we present the KS statistics and their associated p-values. In all cases considered, using any usual significance level, we accept the hypothesis that the data come from gamma distribution with parameters specified above.

\begin{table}[h!]
\centering
\begin{tabular}{c|ccccc}
\hline
Given $N\rightarrow$ & one-day & two-day & three-day & four-day & five-day\\
\hline
KS statistic & 0.0720 &  0.0802 &  0.1002 &  0.1737 & 0.2242\\ 
p-value & 0.1229 &  0.3452 & 0.3377 & 0.2287 &  0.3809\\
\hline
\end{tabular}
\caption{Kolmogorov-Smirnov statistics and their associated p-values for the goodness-of-fit of the conditional distributions of the cumulative log-returns given the durations (one-day, two-day, three-day, four-day and five-day).}\label{conditionalX}
\end{table}

\section{The induced L\'evy process}

 As seen before, the bivariate gamma-geometric distribution is infinitely divisible, therefore, we have that (\ref{idcf}) is a characteristic function for any real $r>0$. This characteristic function is associated with the bivariate random vector
 \begin{eqnarray*}
 (R(r),v(r))=\left(\sum_{i=1}^{T}X_i+G,r+T\right),
 \end{eqnarray*}
where $\{X_i\}_{i=1}^\infty$ are iid random variables following $\Gamma(\alpha,\beta)$ distribution, $G\sim\Gamma(r\alpha,\beta)$, $T$ is a discrete random variable with $\mbox{NB}(r,p)$ distribution and all random variables involved are mutually independent. Hence, it follows that the BGG distribution induces a L\'evy process $\{(X(r),\mbox{NB(r)}),\,\, r\geq0\}$, which has the following stochastic representation:
\begin{eqnarray}\label{flp}
\{(X(r),N(r)),\,\, r\geq0\}\stackrel{d}{=}\left\{\left(\sum_{i=1}^{NB(r)}X_i+G(r),r+\mbox{NB}(r)\right),\,\, r\geq0\right\},
\end{eqnarray}  
where the $X_i$'s are defined as before, $\{G(r),\,\, r\geq0\}$ is a gamma L\'evy process and $\{\mbox{NB}(r),\,\, r\geq0\}$ is a negative binomial L\'evy process, both with characteristic functions given by 
\begin{eqnarray*}
E\left(e^{itG(r)}\right)=\left(\frac{\beta}{\beta-it}\right)^{\alpha r}, \quad t\in\mathbb{R},
\end{eqnarray*}
and
\begin{eqnarray*}
E\left(e^{isN(r)}\right)=\left(\frac{p}{1-(1-p)e^{is}}\right)^r, \quad s\in\mathbb{R},
\end{eqnarray*}
respectively. All random variables and processes involved in (\ref{flp}) are mutually independent.

From the process defined in (\ref{flp}), we may obtain other related L\'evy motions by deleting $r$ and/or $G(r)$. Here, we focus on the L\'evy process given by (\ref{flp}) and by deleting $r$. In this case, we obtain the following stochastic representation for our process:   
\begin{eqnarray}\label{lp}
\{(X(r),\mbox{NB}(r)),\,\, r\geq0\}\stackrel{d}{=}\left\{\left(G(r+\mbox{NB}(r)),\mbox{NB}(r)\right),\,\, r\geq0\right\}.
\end{eqnarray} 
Since both processes (the left and the right ones of the equality in distribution) in (\ref{lp}) are L\'evy, the above result follows by noting that for all fixed $r$, we have $\sum_{i=1}^{NB(r)}X_i+G(r)|\mbox{NB}(r)=k\sim\Gamma(\alpha(r+k),\beta)$. One may also see that the above result follows 
from the stochastic self-similarity property discussed, for example, by Kozubowski and Podg\'orski (2007): a gamma L\'evy
process subordinated to a negative binomial process with drift is again a gamma process. 

The characteristic function corresponding to the (\ref{lp}) is given by
\begin{eqnarray}\label{lpcf}
\Phi^*(t,s)\equiv E\left(e^{itX(r)+isNB(r)}\right)=\left\{\frac{p\beta^\alpha}{(\beta-it)^\alpha-e^{is}\beta^\alpha(1-p)}\right\}^r, 
\end{eqnarray}
for $t,s\in\mathbb{R}$. With this, it easily follows that the characteristic function of the marginal process $\{X(r),\,\, r\geq0\}$ is 
\begin{eqnarray*}\label{lpcfm}
E\left(e^{itX(r)}\right)=\left\{\frac{p\beta^\alpha}{(\beta-it)^\alpha-\beta^\alpha(1-p)}\right\}^r.
\end{eqnarray*}
Since the above characteristic function corresponds to a random variable whose density is an infinite mixture of gamma densities (see Subsection 5.1), we have that $\{X(r),\,\, r\geq0\}$ is an infinite mixture of gamma L\'evy process (with negative binomial weights).
Then, we obtain that the marginal processes of $\{(X(r),\mbox{NB}(r)),\,\, r\geq0\}$ are infinite mixture of gamma and negative binomial processes. Therefore, we define that $\{(X(r),\mbox{NB}(r)),\,\, r\geq0\}$ is a $\mbox{BMixGNB}(\beta,\alpha,p)$ L\'evy process. We notice that, for the choice $\alpha=1$ in (\ref{lp}), we obtain the bivariate process with gamma and negative binomial marginals introduced by Kozubowski et al.\,(2008), named BGNB L\'evy motion. 

As noted by Kozubowski and Podg\'orski (2007), if $\{\widetilde{\mbox{NB}}(r),\,\, r\geq0\}$ is a negative binomial process, with parameter $q\in(0,1)$, independent of another negative binomial process \{$\mbox{NB}(r),\,\, r\geq0\}$ with parameter $p\in(0,1)$, then the changed time process $\{\mbox{NB}^*(r),\,\,r\geq0\}=\{\mbox{NB}(r+\widetilde{\mbox{NB}}(r)),\,\,r\geq0\}$ is a negative binomial process with parameter $p^*=pq/(1-p+pq)$. With this and (\ref{lp}), we have that the changed time process $\{(G(r+\mbox{NB}^*(r)),\mbox{NB}(r+\widetilde{\mbox{NB}}(r))),\,\, r\geq0\}$ is a $\mbox{BMixGNB}(\beta,\alpha,p^*)$ L\'evy process.

In what follows, we derive basic properties of the bivariate distribution of the BMixGNB process for fixed $r>0$ and discuss estimation by maximum likelihood and inference for large sample. From now on, unless otherwise mentioned, we will consider $r>0$ fixed.

\subsection{Basic properties of the bivariate process for fixed $r>0$}

\begin{figure}[h!]
	\centering
		\includegraphics[width=0.33\textwidth]{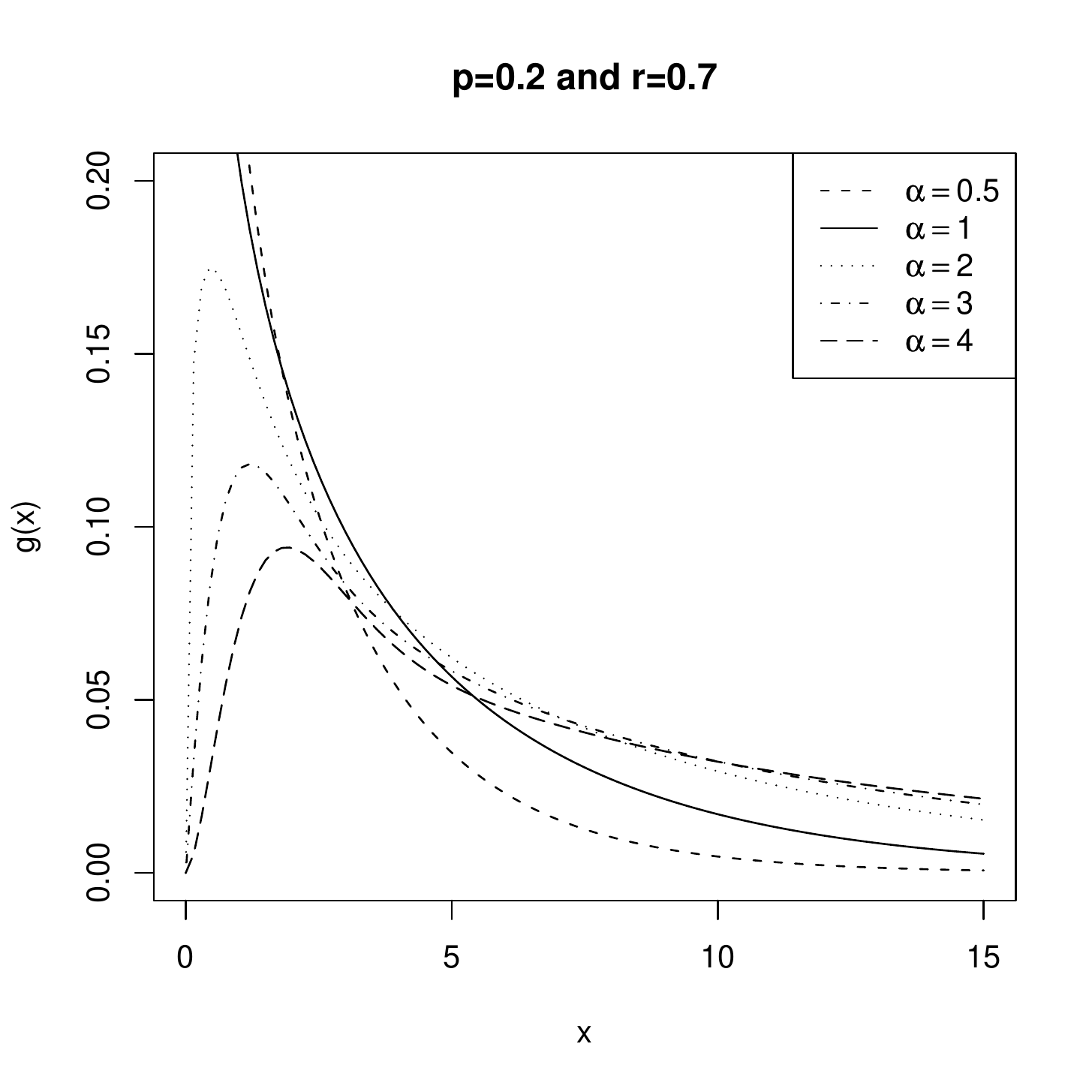}\includegraphics[width=0.33\textwidth]{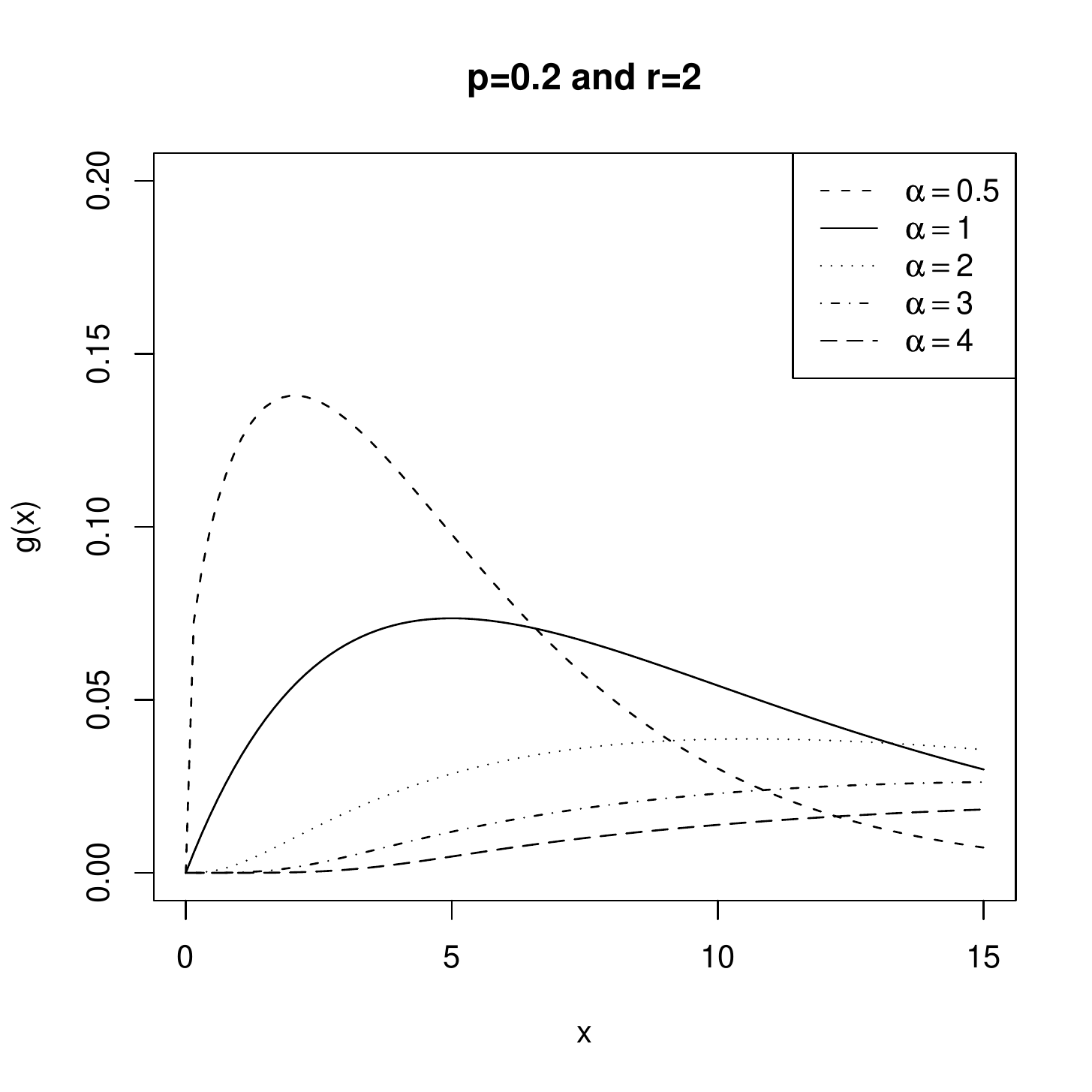}
		\includegraphics[width=0.33\textwidth]{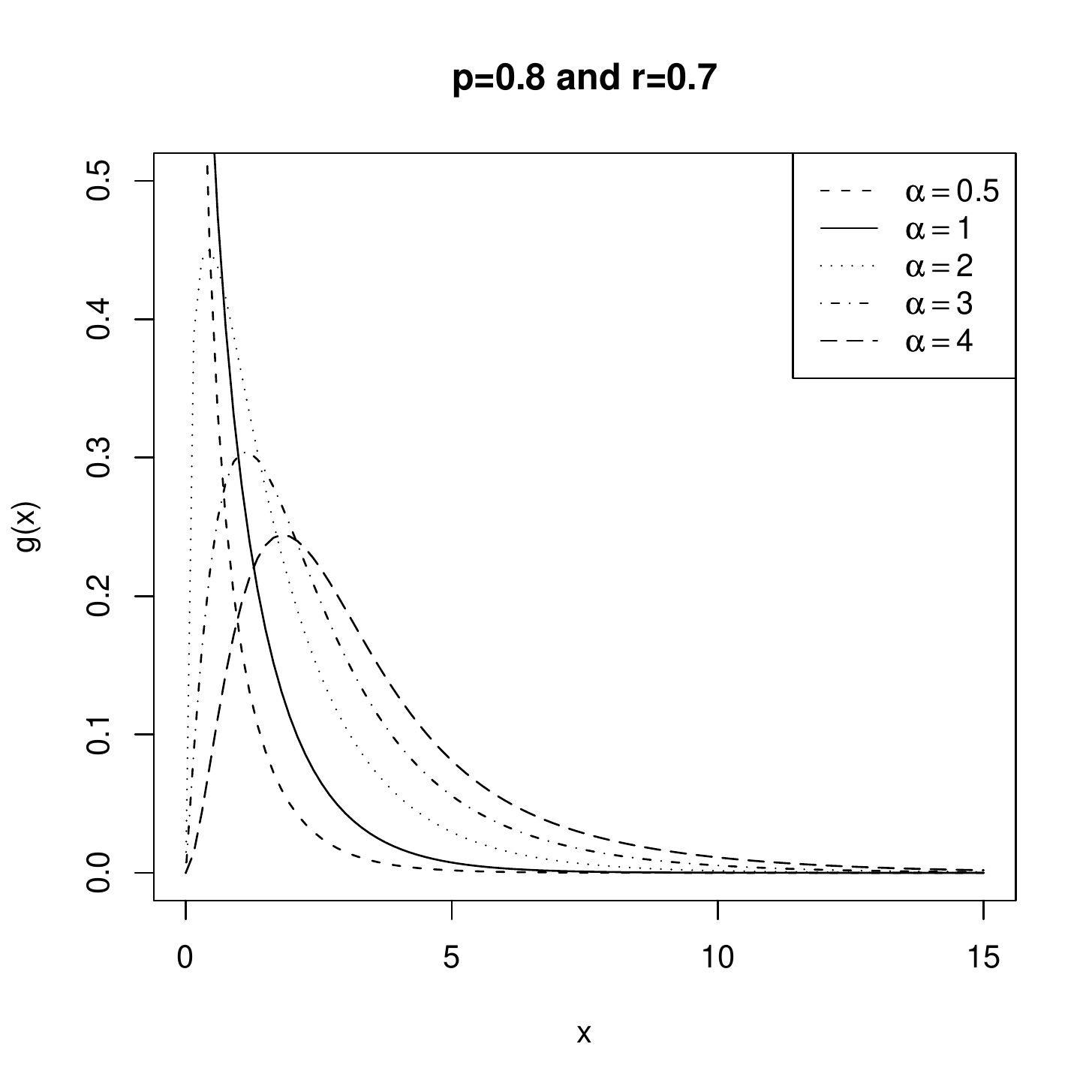}\includegraphics[width=0.33\textwidth]{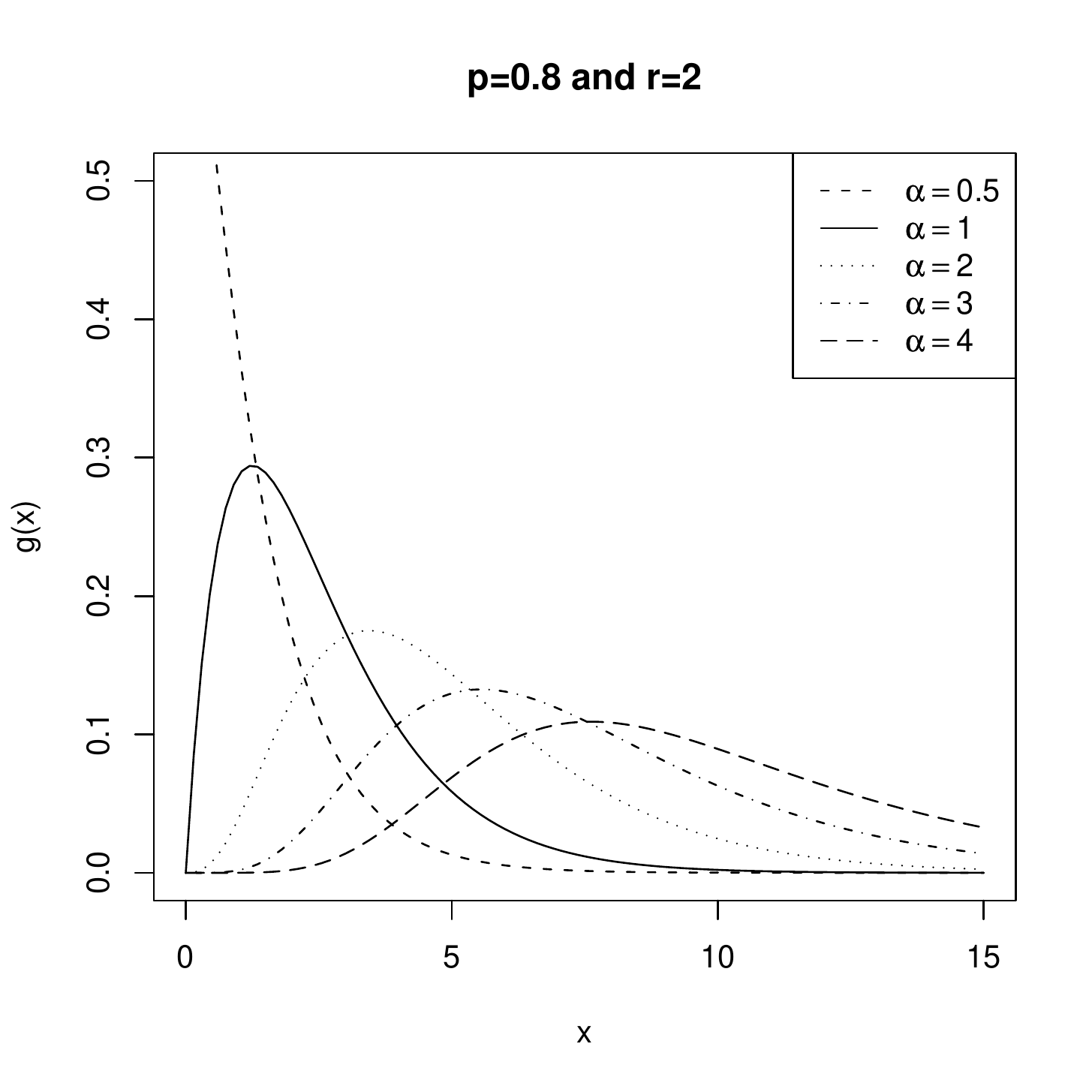}
	\caption{Graphics of the marginal density of $Y$ for $\beta=1$, $\alpha=0.5,1,2,3,4$, $p=0.2,0.8$ and $r=0.7,2$.}
	\label{marginaldensities2}
\end{figure}

For simplicity, we will denote $(Y,M)=(X(r),\mbox{NB}(r))$. From stochastic representation (\ref{lp}), it is easy to see that the joint density and distribution function of $(Y,M)$ are
\begin{eqnarray}\label{pdfr}
g_{Y,M}(y,n)=\frac{\Gamma(n+r)p^r(1-p)^n}{n!\Gamma(r)\Gamma(\alpha(r+n))}\beta^{\alpha(r+n)}y^{\alpha(r+n)-1}e^{-\beta y}
\end{eqnarray} 
and
\begin{eqnarray*}
P(Y\leq y, M\leq n)=\frac{p^r}{\Gamma(r)}\sum_{j=0}^n(1-p)^j\frac{\Gamma(j+r)}{j!\Gamma(\alpha(r+j))}\Gamma_{\beta y}(\alpha(r+j)),
\end{eqnarray*}
for $y>0$ and $n\in\mathbb{N}\cup\{0\}$. Making $\alpha=1$ in (\ref{pdfr}), we obtain the $\mbox{BGNB}$ distribution (bivariate distribution with gamma and negative binomial marginals) as particular case. This model was introduced and studied by Kozubowski et al. (2008). We have that the marginal distribution of $M$ is negative binomial with probability mass function
given in (\ref{nbpf}). The marginal density of $Y$ is given by
\begin{eqnarray*}\label{densityy}
g_Y(y)=\sum_{n=0}^\infty P(M=n)g(y;\alpha(r+n),\beta), \quad y>0,
\end{eqnarray*}
where $g(\cdot;\alpha,\beta)$ is the density of a gamma variable as defined in the Section 2. Therefore, the above density is an infinite mixture of gamma densities (with negative binomial weigths). Since the marginal distributions of $(Y,M)$ are infinite mixture of gamma and negative binomial distributions, we denote $(Y,M)\sim\mbox{BMixGNB}(\beta,\alpha,p,r)$.
Some plots of the marginal density of $Y$ are displayed in the Figure \ref{marginaldensities2}, for $\beta=1$ and some values of $\alpha$, $p$ and $r$.

The conditional distribution of $Y|M=k$ is gamma with parameters $\alpha(r+k)$ and $\beta$, while the conditional probability distribution function of $M|Y=y$ is given by
$$P(M=n|Y=y)=\frac{\Gamma(n+r)}{n!\Gamma(\alpha(n+r))}[(1-p)(\beta y)^\alpha]^n\bigg/\sum_{j=0}^\infty\frac{\Gamma(j+r)}{j!\Gamma(\alpha(j+r))}[(1-p)(\beta y)^\alpha]^j,$$
for $n=0,1,\ldots$, which belongs to one-parameter power series distributions if $\alpha$ and $r$ are known. In this case, the parameter is $(1-p)(\beta y)^\alpha$.  For positive integers $m\leq n$ and real $y>0$, it follows that
$$P(Y\leq y, M\leq m|M\leq n)=\sum_{j=0}^m\frac{\Gamma(j+r)(1-p)^j}{j!\Gamma(\alpha(j+r))}\Gamma_{\beta y}(\alpha(r+j))\bigg/\sum_{j=0}^n\frac{\Gamma(j+r)}{j!}(1-p)^j$$
and for $0<x\leq y$ and positive integer $n$
$$P(Y\leq x, M\leq n|Y\leq y)=\frac{\sum_{j=0}^n\frac{\Gamma(j+r)(1-p)^j}{j!\Gamma(\alpha(j+r))}\Gamma_{\beta x}(\alpha(r+j))}{\sum_{j=0}^\infty\frac{\Gamma(j+r)(1-p)^j}{j!\Gamma(\alpha(j+r))}\Gamma_{\beta y}(\alpha(r+j))}.$$

The moments of a random vector $(Y,M)$ following $\mbox{BMixGNB}(\beta,\alpha,p,r)$ distribution may be obtained by $E(Y^nM^k)=(-i)^{n+k}\partial^{n+k}\Phi^*(t,s)/\partial t^n\partial s^k|_{t,s=0}$, where $\Phi^*(t,s)$ is the characteristic function given in (\ref{lpcf}). It follows that the product moments are given by
\begin{eqnarray}\label{momBMixGNB}
E(Y^nM^k)=\frac{p^r\Gamma(n)}{\beta^n\Gamma(r)}\sum_{m=0}^\infty\frac{m^k(1-p)^m\Gamma(m+r)}{m!B(\alpha(r+m),n)}.
\end{eqnarray}
The covariance matrix of $(Y,M)$ is given by $r\Sigma$, where $\Sigma$ is defined in (\ref{cov}). The correlation coefficient is given by $\rho$, which is defined in the Subsection 2.2. Further, an expression for the $n$th marginal moment of $Y$ may be obtained by taking $k=0$ in (\ref{momBMixGNB}). If $\{W(r),\,r>0\}$ is a $\mbox{BMixGNB}(\beta,\alpha,p)$ L\'evy motion, one may check that $\mbox{cov}(W(t),W(s))=\min(t,s)\Sigma$. 

The $\mbox{BMixGNB}$ law may be represented by a convolution between a bivariate distribution (with gamma and degenerate at 0 marginals) and a compound Poisson distribution. Such a representation is given by
\begin{eqnarray*}
(Y,M)\stackrel{d}{=}(G,0)+\sum_{i=1}^{Q}(G_i,Z_i),
\end{eqnarray*}
with all random variables above defined as in the formula (\ref{cp}), but here we define $G\sim\Gamma(\alpha r,\beta)$ and $\lambda=-r\log p$.
We end this Subsection by noting that if $\{(Y_i,M_i)\}_{i=1}^n$ are independent random vectors with $(Y_i,M_i)\sim\mbox{BMixGNB}(\beta,\alpha,p,r_i)$, then
\begin{eqnarray*}
\sum_{i=1}^n (Y_i,M_i)\sim \mbox{BMixGNB}\left(\beta,\alpha,p,\sum_{i=1}^nr_i\right).
\end{eqnarray*}
One may easily check the above result by using characteristic function (\ref{lpcf}).

\subsection{Estimation and inference for the $\mbox{BMixGNB}$ distribution}

Suppose $(Y_1,M_1), \ldots, (Y_n,M_n)$ is a random sample from $\mbox{BMixGNB}(\beta,\alpha,p,\tau)$ distribution. Here the parameter vector will be denoted by $\theta^\dag=(\beta,\alpha,p,\tau)^\top$. The log-likelihood function, denoted by $\ell^\dag$, is given by
\begin{eqnarray*}
\ell^\dag&\propto& n\{\tau\alpha\log\beta-\log\Gamma(\tau)+\tau\log p\}-n\beta\bar{X}_n+n\{\log(1-p)+\alpha\log\beta\}\bar{M}_n\\&+&\sum_{i=1}^n\log\Gamma(M_i+\tau)-
\sum_{i=1}^n\log\Gamma(\alpha(M_i+\tau))+\alpha\sum_{i=1}^n(M_i+\tau)\log X_i,
\end{eqnarray*}
where $\bar{M}_n=\sum_{i=1}^nM_i/n$. 

The associated score function $U^\dag(\theta^\dag)=(\partial\ell^\dag/\partial\beta,\partial\ell^\dag/\partial\alpha,\partial\ell^\dag/\partial p,\partial\ell^\dag/\partial\tau)$ has its components given by
\begin{eqnarray*}
\frac{\partial\ell^\dag}{\partial\beta}&=&n\left\{\frac{\alpha}{\beta}(\tau+\bar{M}_n)-\bar{X}_n\right\},\\ \frac{\partial\ell^\dag}{\partial\alpha}&=&n(\tau+\bar{M}_n)\log\beta+\sum_{i=1}^n(\tau+M_i)\{\log X_i-\Psi(\alpha(\tau+M_i))\},\\
\frac{\partial\ell^\dag}{\partial p}&=&-\frac{n\bar{M}_n}{1-p}+\frac{n\tau}{p},\\
\frac{\partial\ell^\dag}{\partial\tau}&=&n\left\{\log(p\beta^\alpha)-\Psi(\tau)\right\}+\sum_{i=1}^n\left\{\alpha[\log X_i-\Psi(\alpha(\tau+M_i))]+\Psi(\tau+M_i)\right\}.
\end{eqnarray*}

Hence, the maximum likelihood estimators of $\beta$ and $p$ are respectively given by 
\begin{eqnarray}\label{mles2}
\widehat\beta=\widehat\alpha\frac{\widehat\tau+\bar{M}_n}{\bar{X}_n}\quad \mbox{and} \quad \widehat{p}=\frac{\widehat\tau}{\widehat\tau+\bar{M}_n},
\end{eqnarray}
while the maximum likelihood estimators of $\alpha$ and $\tau$ are found by solving the nonlinear system of equations
\begin{eqnarray*}
n(\widehat\tau+\bar{M}_n)\log\left(\widehat\alpha\frac{\widehat\tau+\bar{M}_n}{\bar{X}_n}\right)+\sum_{i=1}^n(\widehat\tau+M_i)\{\log X_i-\Psi(\widehat\alpha(\widehat\tau+M_i))\}=0
\end{eqnarray*}
and
\begin{eqnarray}\label{mletau}
\widehat\alpha\left\{n\log\left(\widehat\alpha\frac{\widehat\tau+\bar{M}_n}{\bar{X}_n}\right)+\sum_{i=1}^n\left\{\log X_i-\Psi(\widehat\alpha(\widehat\tau+M_i))\right\}\right\}=\nonumber\\n\left\{\Psi(\widehat\tau)-\log\left(\frac{\widehat\tau}{\widehat\tau+\bar{M}_n}\right)\right\}-\sum_{i=1}^n\Psi(\widehat\tau+M_i).
\end{eqnarray}
After some algebra, we obtain that Fisher's information matrix is
\begin{eqnarray*}
J^\dag(\theta^\dag)=
\left(\begin{array}{llll}
\kappa^\dag_{\beta\beta} & \kappa^\dag_{\beta\alpha} & 0 & \kappa^\dag_{\beta\tau}\\
\kappa^\dag_{\beta\alpha} & \kappa^\dag_{\alpha\alpha} & 0 & \kappa^\dag_{\alpha\tau}\\
0 & 0 & \kappa^\dag_{pp} & \kappa^\dag_{p\tau} \\
 \kappa^\dag_{\beta\tau} &  \kappa^\dag_{\alpha\tau} &  \kappa^\dag_{p\tau}&  \kappa^\dag_{\tau\tau}\\
\end{array}\right),
\end{eqnarray*}
with 
\begin{eqnarray*}
&&\kappa^\dag_{\beta\beta}=\frac{\alpha\tau}{\beta^2p},\quad \kappa^\dag_{\beta\alpha}=-\frac{\tau}{p\beta},\quad\kappa^\dag_{\beta\tau}=-\frac{\alpha}{\beta},\\ &&\kappa^\dag_{\alpha\alpha}=\frac{p^\tau}{\Gamma(\tau)}\sum_{j=0}^\infty (\tau+j)^2(1-p)^j\Psi'(\alpha(\tau+j))\frac{\Gamma(\tau+j)}{j!}, \quad \kappa^\dag_{pp}=\frac{\tau}{p^2(1-p)},\\
&&\kappa^\dag_{\alpha\tau}=\frac{\alpha p^\tau}{\Gamma(\tau)}\sum_{j=0}^\infty (1-p)^j(\tau+j)\Psi'(\alpha(\tau+j))\frac{\Gamma(\tau+j)}{j!},
\quad\kappa^\dag_{p\tau}=-\frac{1}{p},\\ 
&&\kappa^\dag_{\tau\tau}=\Psi'(\tau)+\frac{p^\tau}{\Gamma(\tau)}\sum_{j=0}^\infty (1-p)^j\{\alpha^2\Psi'(\alpha(\tau+j))-\Psi'(\tau+j)\}\frac{\Gamma(\tau+j)}{j!}.
\end{eqnarray*}
So we obtain that the asymptotic distribution of $\sqrt{n}(\widehat\theta^\dag-\theta^\dag)$ is trivariate normal with null mean and covariance matrix $J^{\dag\,-1}(\theta^\dag)$, where $J^{\dag\,-1}(\cdot)$ is the inverse of the information matrix $J^\dag(\cdot)$ defined above. The likelihood ratio, Wald and Score tests may be performed in order to test the hypotheses $H_0 \mbox{:} \,\,\alpha=1$ versus $H_1 \mbox{:} \,\,\alpha\neq1$, that is, to compare $\mbox{BGNB}$ and $\mbox{BMixGNB}$ fits. Further, we may test the $\mbox{BMixGNB}$ model versus the BGG or BEG models, which corresponds to the null hypotheses $H_0 \mbox{:} \,\,\tau=1$ and $H_0 \mbox{:} \,\,\alpha=\tau=1$, respectively. 

As made in the Subsection 4.2, we here propose the reparametrization $\mu=\alpha/\beta$. We now denote the parameter vector by $\theta^{\star}=(\mu,\alpha,p,\tau)^\top$.  With this, one may check that the MLEs of $p$ and $\mu$ are given by (\ref{mles2}) and $\widehat\mu=\bar{X}_n/(\widehat\tau+\bar{M}_n)$. The MLEs of $\tau$ and $\alpha$ are obtained by solving the nonlinear system of equations (\ref{mletau}) and 
\begin{eqnarray*}
n(\widehat\tau+\bar{M}_n)\left\{\log\left(\widehat\alpha\frac{\widehat\tau+\bar{M}_n}{\bar{X}_n}\right)-\frac{\widehat\tau+\bar{M}_n}{\bar{X}_n}\right\}+\sum_{i=1}^n(\widehat\tau+M_i)\{\log X_i-\Psi(\widehat\alpha(\widehat\tau+M_i))\}=0.
\end{eqnarray*} 
Under this proposed reparametrization, the Fisher's information matrix becomes
\begin{eqnarray*}
J^\star(\theta^{\star} )=
\left(\begin{array}{llll}
\kappa^\star_{\mu\mu} & 0 & 0 & \kappa^\star_{\mu\tau}\\
0 & \kappa^\star_{\alpha\alpha} & 0 & \kappa^\star_{\alpha\tau}\\
0 & 0 & \kappa^\star_{pp} & \kappa^\star_{p\tau} \\
 \kappa^\star_{\mu\tau} &  \kappa^\star_{\alpha\tau} &  \kappa^\star_{p\tau}&  \kappa^\star_{\tau\tau}\\
\end{array}\right),
\end{eqnarray*}
where its elements are given by 
\begin{eqnarray*}
&&\kappa^\star_{\mu\mu}=\frac{\alpha\tau}{\mu^2p},\quad \kappa^\star_{\mu\tau}=\frac{\alpha}{\mu}, \quad\kappa^\star_{\alpha\alpha}=\frac{p^\tau}{\Gamma(\tau)}\sum_{j=0}^\infty (1-p)^j(\tau+j)^2\Psi'(\alpha(\tau+j))\frac{\Gamma(\tau+j)}{j!}-\frac{\tau}{\alpha p},\\
&&\kappa^\star_{\alpha\tau}=\frac{\alpha p^\tau}{\Gamma(\tau)}\sum_{j=0}^\infty (1-p)^j(\tau+j)\Psi'(\alpha(\tau+j))\frac{\Gamma(\tau+j)}{j!}-1,
\quad \kappa^\star_{pp}=\kappa^\dag_{pp},\\
&& \kappa^\star_{p\tau}=\kappa^\dag_{p\tau}\quad\mbox{and}\quad \kappa^\star_{\tau\tau}=\kappa^\dag_{\tau\tau}.
\end{eqnarray*}

We have that $\kappa^\star_{\mu\alpha}=0$, that is, $\mu$ and $\alpha$ are orthogonal parameters in contrast with the parameters $\beta$ and $\alpha$ considered previously, where $\kappa^\dag_{\beta\alpha}\neq0$. Further, we have that $\sqrt{n}(\widehat\theta^\star-\theta^\star)\rightarrow N_4(0,J^{\star\,-1}(\theta^\star))$ as $n\rightarrow\infty$, where the covariance matrix $J^{\star\,-1}(\theta^\star)$ is the inverse of the information matrix $J^\star(\theta^\star)$.

\section{Concluding remarks}

We introduced and studied the bivariate gamma-geometric (BGG) law, which extends the bivariate exponential-geometric (BEG) law proposed by Kozubowski and Panorska (2005). The marginals of our model are infinite mixture of gamma and geometric distributions. Several results and properties were obtained such as joint density and survival functions, conditional distributions, moment generation and characteristic functions, product moments, covariance matrix, geometric stability and stochastic representations. 

We discussed estimation by maximum likelihood and inference for large sample. Further, a reparametrization was suggested in order to obtain orthogonality of the parameters. An application to exchange rates between Brazilian real and U.K. pounds, quoted in Brazilian real, was presented. There our aim was to model jointly the magnitude and duration of the consecutive positive log-returns. In that application, we showed that the BGG model and its marginal and conditional distributions fitted suitably the real data set considered. Further, we performed the likelihood ratio and Wald tests and  both rejected (with significance level at 5\%) the hypothesis that the data come from BEG distribution in favor of the BGG distribution. 

We showed that our bivariate law is infinitely divisible and, therefore, induces a L\'evy process, named $\mbox{BMixGNB}$ L\'evy motion. We also derived some properties and results of this process, including a study of its distribution at fixed time. Our proposed L\'evy motion has infinite mixture of gamma and negative binomial marginal processes and generalizes the one proposed by Kozubowski et al. (2008), whose marginals are gamma and negative binomial processes. Estimation and inference for the parameters of the distribution of our process at fixed time were also discussed, including a reparametrization to obtain a partial orthogonality of the parameters.

\section*{Acknowledgements}

\noindent I thank the anonymous referee for their careful reading, comments and suggestions. I also gratefully acknowledge financial support from {\it Conselho Nacional de Desenvolvimento Cient\'ifico e Tecnol\'ogico} (CNPq-Brazil).

\end{document}